\documentclass[conference]{IEEEtran}
\IEEEoverridecommandlockouts
\UseRawInputEncoding
\usepackage{cite}
\usepackage{amsmath,amssymb,amsfonts}
\usepackage{algorithmic}
\usepackage{graphicx}
\usepackage{textcomp}
\usepackage{xcolor}
\usepackage{threeparttable}
\usepackage[linesnumbered,boxed,ruled,commentsnumbered]{algorithm2e}
\usepackage{bm}
\usepackage{amsfonts}
\usepackage{pifont}
\usepackage{booktabs}
\usepackage{subfigure}
\usepackage{float}

\def\BibTeX{{\rm B\kern-.05em{\sc i\kern-.025em b}\kern-.08em
    T\kern-.1667em\lower.7ex\hbox{E}\kern-.125emX}}

\newtheorem{theorem}{\bf Theorem}  [section]

\newtheorem{assumption}{\bf Assumption}[section]
\newtheorem{definition}{\bf Definition}[section]

\begin{document}

\title{Proof of User Similarity: the Spatial Measurer of Blockchain\\
\author{Shengling~Wang,~\IEEEmembership{Senior Member,~IEEE,}
        Lina~Shi,
        Hongwei~Shi,
        Yifang~Zhang,
        Qin~Hu,\\
        and Xiuzhen~Cheng,~\IEEEmembership{Fellow,~IEEE}
\thanks{Shengling Wang is with the School of Artificial Intelligence and the Faculty of Education, Beijing Normal University, Beijing, China. E-mail: wangshengling@bnu.edu.cn.}
\thanks{Lina Shi, Hongwe Shi (Corresponding author) and Yifang Zhang are with the School of Artificial Intelligence, Beijing Normal University, Beijing, China. E-mail: 201821210031@mail.bnu.edu.cn, hongweishi@mail.bnu.edu.cn, zyfyydgq@mail.bnu.edu.cn.}
\thanks{Qin Hu is with the Department of Computer and Information Science, Indiana University - Purdue University Indianapolis, IN, USA. E-mail:qinhu@iu.edu.}
\thanks{Xiuzhen Cheng is with the School of Computer Science and Technology, Shandong University (SDU), Shandong, China. E-mail: xzcheng@sdu.edu.cn.}
}
}

\maketitle

\begin{abstract}
Although proof of work (PoW) consensus dominates the current blockchain-based systems mostly, it has always been criticized for the uneconomic brute-force calculation. As alternatives, energy-conservation and energy-recycling mechanisms heaved in sight. In this paper, we propose {\it proof of user similarity} (PoUS), a distinct energy-recycling consensus mechanism, harnessing the valuable computing power to calculate the similarities of users, and enact the calculation results into the packing rule. However, the expensive calculation required in PoUS challenges miners in participating, and may induce plagiarism and lying risks. To resolve these issues, PoUS embraces the {\it best-effort} schema by allowing miners to compute partially. Besides, a voting mechanism based on the two-parties computation and Bayesian truth serum is proposed to guarantee privacy-preserved voting and truthful reports. Noticeably, PoUS distinguishes itself in recycling the computing power {\it back to blockchain} since it turns the resource wastage to facilitate refined cohort analysis of users, serving as the {\it spatial measurer} and enabling a {\it searchable} blockchain. 
We build a prototype of PoUS and compare its performance with PoW. The results show that PoUS outperforms PoW in achieving an average TPS improvement of 24.01\% and an average confirmation latency reduction of 43.64\%. Besides, PoUS functions well in mirroring the spatial information of users, with negligible computation time and communication cost.


\end{abstract}


\section{Introduction}\label{Sec-introduction}
\IEEEPARstart{B}{lockchain} \cite{bitcoin} has symbolized a thrilling breakthrough that can bootstrap confidence among distrustful parties on an Internet scale. It is essentially a linear chain of blocks, with each being appended in chronological order via {\it consensus}. The consensus, therefore, characterizes the sequentiality of blocks from the {\it temporal scale}, standing as the main pillar for system security. Among the existing consensus protocols, PoW dominates the current blockchain-based systems mostly, where the participants (called {\it miners}) struggle to figure out computational puzzles by brute-force calculation (called {\it mining}), in order to get the accounting right and obtain the reward. Despite its security and robustness, PoW is often criticized for its uneconomical feature, thus striking hot debates on whether other mechanisms are available to function better than PoW.

Pioneer countermeasures are the {\it energy-conservation} and {\it energy-recycling} alternatives. In detail, the former mechanisms dwindle energy consumption through exerting other selection criteria, such as stakes  (proof of stake \cite{pos}) and votes (delegated proof of stake \cite{eos}). Unfairness is the main drawback of these schemes because the intrinsic capacity bias may cause the phenomenon of {\it the rich get richer} \cite{PoFL,epos}. This alarms us that sacrificing expensive resources to achieve a fair and secure consensus is necessary and inevitable, thus inspiring the {\it energy-recycling} mechanisms, which repurpose the computing power investments into other useful works.

The ongoing energy-recycling consensus mechanisms devote costly computing power to working out problems in other fields other than blockchain itself, like mathematics \cite{primecoin}, database \cite{permacoin}, and machine learning \cite{PoFL}, \cite{du}. In this paper, we propose {\it proof of user similarity} (PoUS), where we rearrange the computing power to calculate the similarities of users who issue transactions in blockchain, and enact the calculation results into the packing rule. Besides our PoUS can inherit the essence of energy-recycling consensus mechanisms, it is advocated from the following aspects:
\begin{itemize}
  \item First, PoUS reinvests the valuable computing power back to blockchain instead of contributing to other fields. Motivated by {\it keeping the goodies within the family}, the inherent demand of blockchain can be deeply explored, which may energize a more prosperous blockchain.
  \item Second, the similarity-in-design PoUS facilitates {\it cohort analysis} of users \cite{cohort}, which can disclose the population distribution and behavior patterns of users spatially and pave the way for investigating user life cycles and value retentions. In doing so, the consensus in blockchain can be complementarily replenished besides the temporal scale from a fresh angle of {\it spatial dimension}.
  \item Third, transaction packaging based on user similarity enables a {\it searchable} blockchain. Equipped with powerful storage and index technologies, an effective query and retrieval database based on blockchain can be realized.
\end{itemize}



Despite the above merits of PoUS, we encounter the following three challenges when fulfilling it: 1) {\it supply-demand contradiction}. In fact, the user similarity calculation will exhaust a large amount of computing and storage resources, making it impractical for a single miner to carry out the large-scale calculation. However, a fair and secure consensus mechanism calls for involvement jointly; 2) {\it plagiarism risk}. Since no explicit standard exists to measure the calculation result of user similarity, we resort to the voting mechanism to select the most qualified one. However, the transmission of the calculation result of each candidate may provoke plagiarizers who pretend to vote while copying, ruining the fairness of consensus; 3) {\it lying risk}. Despite democracy, the voting mechanism can be easily destroyed by untruthful reports since the real intentions of voters are private information that can be easily hidden. This may lead to interest-oriented liars, who grant their polls to the uncertified candidates, endangering blockchain consequently.

Our work intends to resolve the above challenges to make PoUS practically applicable. To the best of our knowledge, our paper is the first work to employ user similarity calculation as the proof of work for consensus in blockchain. As such, a general framework of PoUS is introduced, where the above challenges are addressed as follows:
\begin{itemize}
  \item To resolve the first challenge, we embrace the {\it best-effort} schema for design. Concretely, our PoUS allows miners to participate in consensus via calculating  user similarity {\it partially}. That is to say, miners are only required to dedicate their resources within their capabilities. As long as the majority of miners operate honestly, large-scale user similarity calculation can be achieved. 
  \item As for the second challenge, a voting mechanism based on two-parties computation (2PC) is presented to leverage the cryptographic primitives to assure correct voting without disclosing any private information of the candidates. By doing this, the plagiarism risk can be well-repelled since the authentic calculation results of each candidate are masked, and more significantly, the accuracy of voting will not be sacrificed.
  \item The third challenge can be well addressed through a Bayesian truth serum-based incentive mechanism. This mechanism can reward a truth-teller more than a liar, thus encouraging interest-driven voters to honestly report their real beliefs. Under this case, the lying risk can be decreased, which facilitates a democratic and secure voting mechanism accordingly.
\end{itemize}

We implement a prototype for PoUS in Python based on BlockSim\footnote{https://github.com/maher243/BlockSim}, 
and compare its performance with PoW. Through extensive simulations, we can conclude that PoUS outperforms PoW in achieving an average TPS improvement of 24.01\% and an average confirmation latency reduction of 43.64\%. Besides, PoUS functions well in reflecting the spatial information of users with negligible computation time and communication cost, distinguishing it as the spatial measurer of blockchain.

The remaining part of our paper is organized as follows. We first introduce PoUS from the top level in Section \ref{Sec-overview}, which proceeds sequentially from mining, voting to packing. The first stage, i.e., user similarity calculation-based mining,  is detailedly described in Section \ref{Sec-mining}. After that, the second stage, i.e., the plagiarism- and lying-proof voting mechanism based on 2PC and Bayesian truth serum, is presented in Section \ref{Sec-voting}. The third stage is demonstrated in Section \ref{Sec-packing}, which expresses the clustering-based packing scheme. We carry out theoretical analysis and develop a prototype in Section \ref{Sec-eva} to give a thorough evaluation of PoUS. Some questions about PoUS are raised and answered in Section \ref{sec-concerns}. Section \ref{Sec-rel} summarizes the related work and finally, Section \ref{Sec-conclusion} concludes our paper.

\section{The Overview of PoUS}\label{Sec-overview}
\begin{figure}
\centering
\includegraphics[scale=0.3]{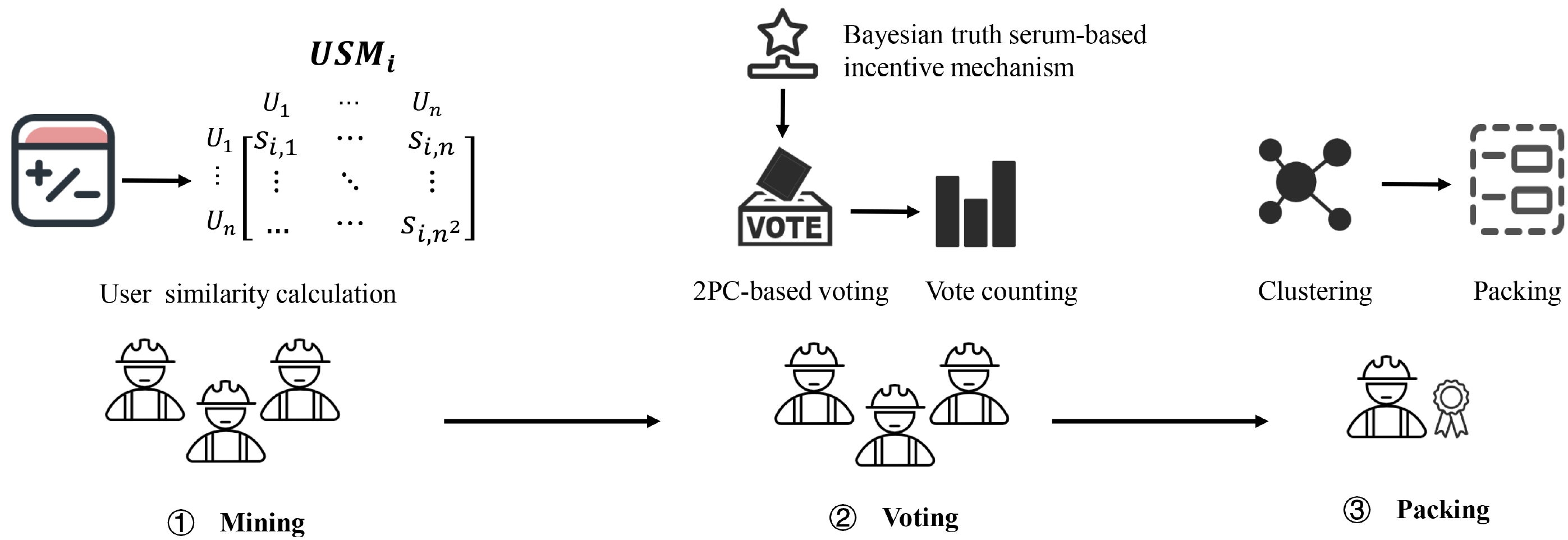}
\caption{The workflow of PoUS consensus.}
\label{ThreeStage}
\end{figure}

At a high level, PoUS involves three stages to reach consensus, which are mining, voting and packing. As shown in Fig. \ref{ThreeStage}, these three stages are proceeded sequentially, with each serving for different purposes. To illustrate:

\ding{172} Mining: We denote the sets of users and miners as $\{U_1, U_2, ..., U_n\}$ and $\{M_1, M_2, ..., M_m\}$, where $n$ and $m$ respectively represent the number of users and miners. In the mining stage, every miner $i\in\{1,2,...,m\}$ conducts user similarity calculation according to its view of current transaction data, and accordingly, obtains the user similarity matrix $\it{USM}_i$, in which each element $s_{i,j}$ represents the similarity between the row and column users and $j\in\{1,2,...,n^2\}$. In particular, the only way for a success-hungry miner to be nominated as the leader\footnote{By leader, we mean the one who gets the accounting right currently.} is to compute its $\it{USM}$ as accurately as possible, which requires extremely huge storage and computing resources. Hence, such a calculation process can be deemed as the mining period as PoW.

\ding{173} Voting: After computing $\it{USM}$s, we resort to the voting mechanism to pick up the most qualified miner as the leader. However, a naive voting mechanism may suffer from the following malfeasances:

\begin{itemize}
  \item {\textbf{plagiarism.}} Intuitively, the candidates should send their ${\it USM}$s to other voters, based on which, the voters can compare to choose the most competent one. However, publishing ${\it USM}$s in the form of clear text may incur plagiarists who embezzle others' computing results. To concur with this, we design a 2PC-based voting mechanism empowered by Garbled Circuits (GC) \cite{GC} and Oblivious Transfer (OT) protocol \cite{OT}, which allows miners to vote without knowing the plain ${\it USM}$s of the candidates, wiping out plagiarists consequently.
  \item {\textbf{lying.}} Although the voting ecology seems democratic, it can be easily ruined by untruthful elections since the real intentions of voters are private information that can be easily hidden. Considering this, we adopt the Bayesian truth serum-based incentive mechanism to elicit truthful votes among miners, which teaches the voters that only if it reports honestly, can it obtain the highest payoff.
\end{itemize}

When all of the miners cast their true votes (no lying) based on their calculated ${\it USM}$s (no plagiarism) according to our voting mechanism, they are required to submit their votes to a vote-counting committee by running a smart contract. After that, the committee can achieve consensus on electing the highest-voted miner as the leader.

\ding{174} Packing: In the last stage of PoUS, the leader first clusters transactions with the best global user similarity, and then bails them into a block according to their priorities, which consider user similarity, transaction fee and waiting time. In light of this, PoUS can portray user distribution from the spatial angle without compromising system performance.

\begin{figure}
\centering
\includegraphics[scale=0.7]{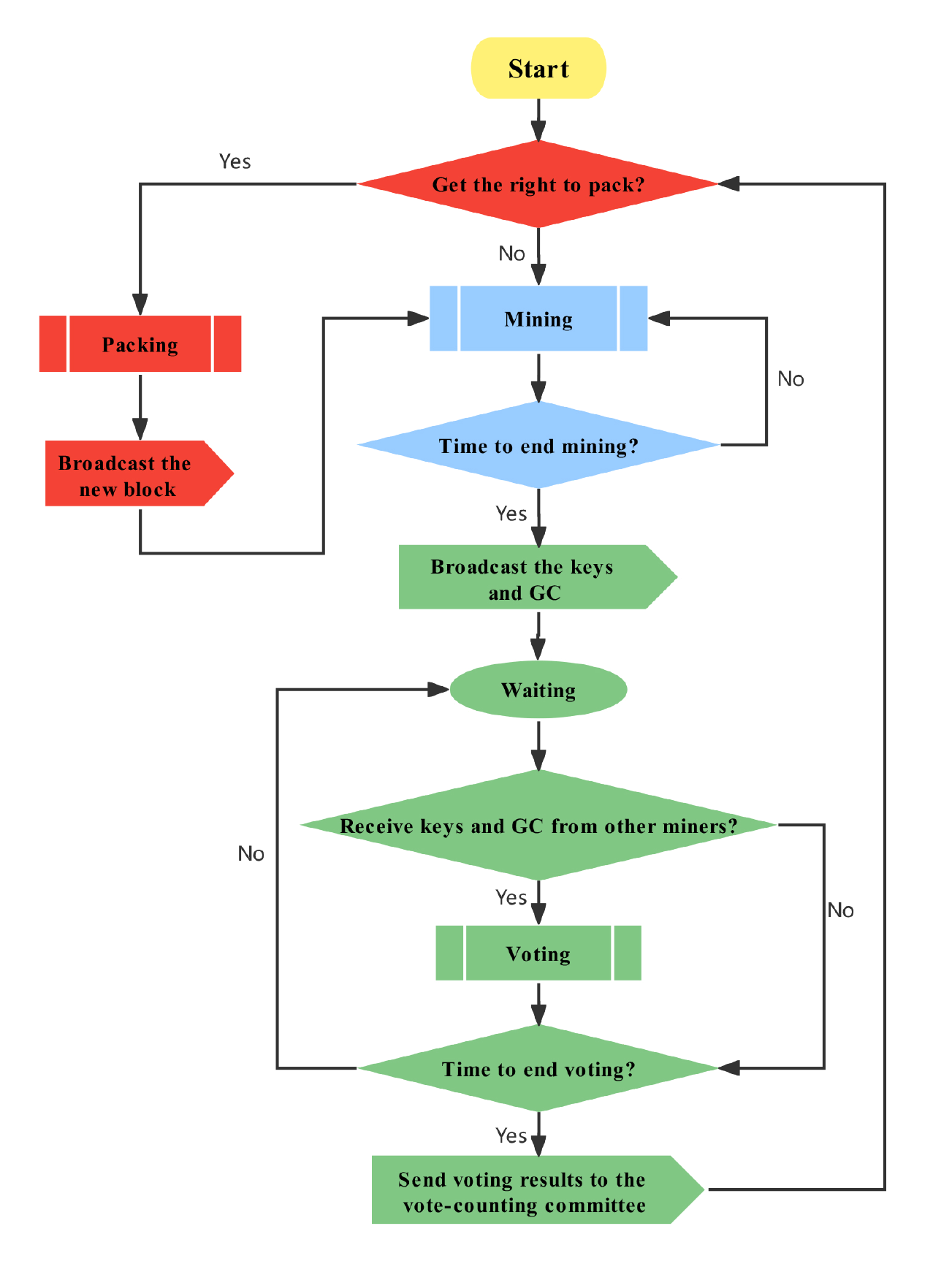}
\caption{The process of PoUS consensus.}
\label{consensus process}
\end{figure}

Fig. \ref{consensus process} demonstrates the above consensus process more elaborately from the miner's perspective, where the mining, voting and packing periods are respectively depicted as the blue, green and red boxes. To begin with, the miner should determine whether it gets the packing right for the current round or not. If yes, it needs to generate a new block by packing transactions based on their priorities and then broadcasting it to the committee for verification, before starting a new round of mining. Otherwise, it can mine directly via calculating its $\it{USM}$ until the time for mining is over. The voting process begins with miners propagating the encrypted keys corresponding to their $\it{USM}$s as well as GCs to each other. When the miner receives the keys and GCs from others, it will then run the OT protocol for completing voting. The above process will continue until the voting time is over, after which the voting results will be sent to the vote-counting committee via running a smart contract. If the miner is nominated as the current leader, it will be informed by the committee, which will also receive the global best user similarity results. This indicates a new round of mining may start right away.



\section{User Similarity Calculation-Based Mining}\label{Sec-mining}
The mining stage based on user similarity calculation will be interpreted detailedly as follows. The miners start to mine when they find out they are not the leaders for the current round, or when they have completed the job of generating a new block if they are the leaders at present. As stated above, the main task of mining is to calculate user similarity accurately. To achieve this, miners should access extensive user data, and convert data into user vectors with well-characterized interests and preferences. After that, miners can conduct similarity measurements between these vectors, which finally leads to $\it{USM}$.

1) {\textbf{User data acquisition.}} The accessibility of massive user data is the cornerstone for accurate user similarity calculation. PoUS sets miners to mine based on both the {\it {historical}} and {\it {latest}} transaction data, where the former denotes the data in the latest $\eta$ blocks of blockchain and the latter represents transactions in the mempool of each miner. These two kinds of data are chosen to span the users' interests and preferences both previously and currently, which can reflect users' behaviors more comprehensively. It is worth noting that different miners may perceive different views of user data, which is mainly caused by network delay. However, we convince that such differences among miners will not bring in conflicts of mining since as long as the majority of miners in the network receive consistent data, the user similarity matrix recognized by most miners can be achieved through the voting process.

2) {\textbf{User vector construction.}} In this period, miners need to convert the obtained user data into vectors for similarity calculation. Multiple vectorization methods are available, and in this paper, we present a simple ``user-type" one as an illustration, where each element in the vector denotes the amount of data in the corresponding category of that user. For instance, in a financial blockchain system, there may be four types of transactions $\it\{Tr_1,Tr_2,Tr_3,Tr_4\}$, and $U_j$ has $3$ pieces of $\it{Tr_1}$ data and $4$ pieces of $\it{Tr_2}$ data, then we can use a $4$-dimensional tuple $(3,4,0,0)$ to characterize $U_j$. In this way, a user-type matrix can be obtained with each user's vector combined. It is worth noting that there is no specific restriction on how to generate user vectors, as long as all the miners act uniformly.

\begin{figure}
\centering
\includegraphics[scale=0.38]{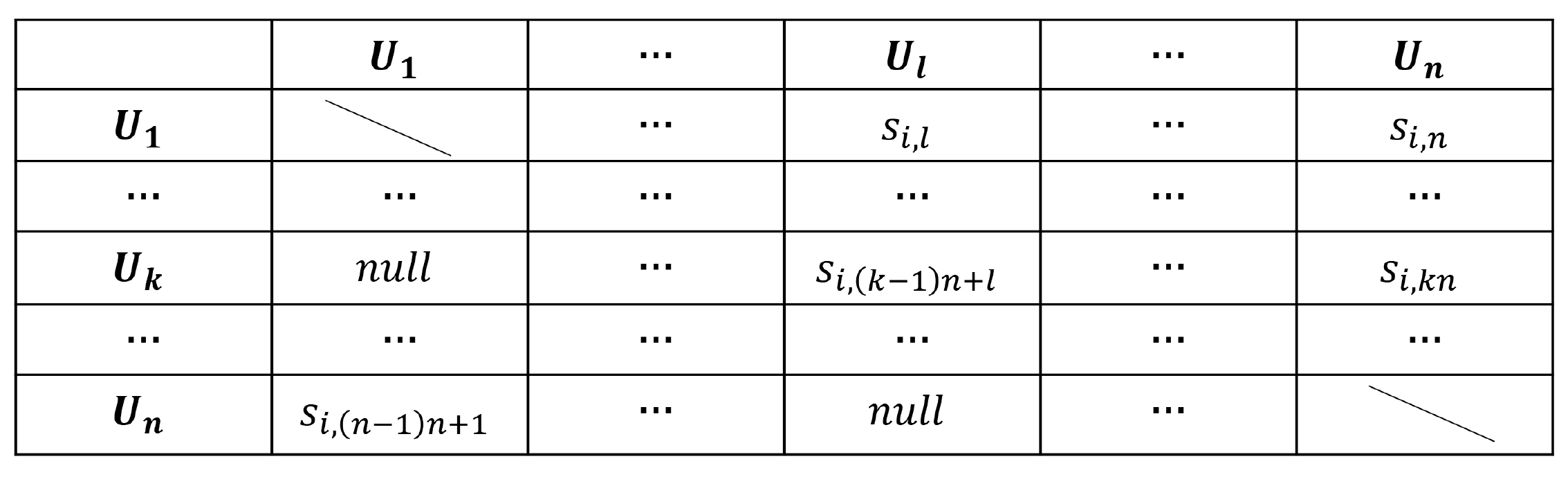}
\caption{The user similarity matrix of $U_i$.}
\label{fig:tu1}
\end{figure}

3) {\textbf{Similarity measurement.}} After constructing user vectors, miners can carry out the similarity measurement through calculating the distance between every pair of them, resulting in the $\it{USM}$. 
All the miners should utilize the same similarity measurement scheme no matter which scheme is specifically employed. However, forcing every miner to compute similarities among all users is impractical because their resources are quite limited. Considering this, the best-effort schema in network service is endorsed in PoUS, so that miners are allowed to submit partial calculation results. We claim that as long as the majority of miners execute correctly, large-scale user similarity calculation can be reached. The $\it{USM}$ of miner $i$, $\it{USM}_i$, is shown in Fig. \ref{fig:tu1}, where the first row and column represent users, and $s_{i,j}$ is the similarity between  users $k$ and $l$ with $j=(k-1)n+l$. Note that for some $\it{USM}s$, there may exist {\it {null}} values because the inadequate capability of computing or storage may cause uncompleted calculation tasks when the mining time ends. 

4) {\textbf{User similarity update.}} Whenever a new transaction arrives, the user vector related to the corresponding users will change, so does the similarity matrix. 
There are two kinds of methods for updating user similarity, which are the recalculation schemes and incremental updating ones\cite{Incremental-1,Incremental-2}. 
No matter which kind of updating method is employed, our PoUS mechanism can function correctly as long as all the miners obey it consistently.

\section{Plagiarism- and Lying-Proof Voting Mechanism}\label{Sec-voting}
In PoUS, there is no measurable standard to select the highest-quality user, so we resort to the voting mechanism. As the miners are allowed to submit partial calculation results, they are also devised to vote partially, which means that miners are only required to focus on the data they have calculated during mining while neglecting others. Take miner $r$ (the voter) votes on miner $i$ (the candidate) as an example. Suppose $M_r$ only computes similarities among $U_1$, $U_2$ and $U_3$, then $M_r$ is merely enforced to compare each valid element in ${\it{USM}}_r$ (i.e., $s_{r,2}$, $s_{r,3}$, etc.) with the same indexed ones in ${\it USM_i}$. Based on the comparison result, $M_r$ votes $1$ or $0$ for each similarity calculated by $M_i$, where $1$ (means approval) implies the difference between the results of $M_r$ and $M_i$ is within a threshold $\theta$ and $0$ (denotes disapproval or abstention) represents the difference is beyond $\theta$ or $M_r$ quits to vote. 
Note that the voting result may be a sparse matrix, hence powerful  data compression protocols \cite{sparse-matrix-1,sparse-matrix-2} which could greatly enhance storage and transmission costs can be applied if necessary.



As mentioned above, a naive voting mechanism may provide a breeding ground for plagiarists or liars, who copy the published ${\it USM}$s from others or collude for not telling the true votes. These two malicious behaviors are meant to be completely hindered in the sense that they undermine the main pillars of consensus, which are fairness and security. In the following, we will introduce the {\it 2PC-based voting mechanism} and the {\it Bayesian truth serum-based incentive mechanism} with each suppressing plagiarists and liars, to fulfill the goal.

\subsection{2PC-Based Voting Mechanism}\label{Section-encryption}
This risk of being copied lies in the fact that each candidate needs to broadcast its calculated ${\it USM}$s to the voters. However, propagating ${\it USM}$s transparently makes room for vicious stealers since everyone can receive and reuse them. Such an issue can be resolved by the secure two-parties computation (2PC) \cite{yao, GC} framework. Essentially, 2PC allows two parties to jointly evaluate the result of the public function $f(x,y)$ without disclosing information about their private input data $x$ and $y$. In this way, voters can judge whether the data to be voted is sufficiently close to their own calculated data (i.e., the difference is within $\theta$) and complete the voting procedure dispensing with privacy leakage.

\begin{figure}
\centering
\includegraphics[scale=0.3]{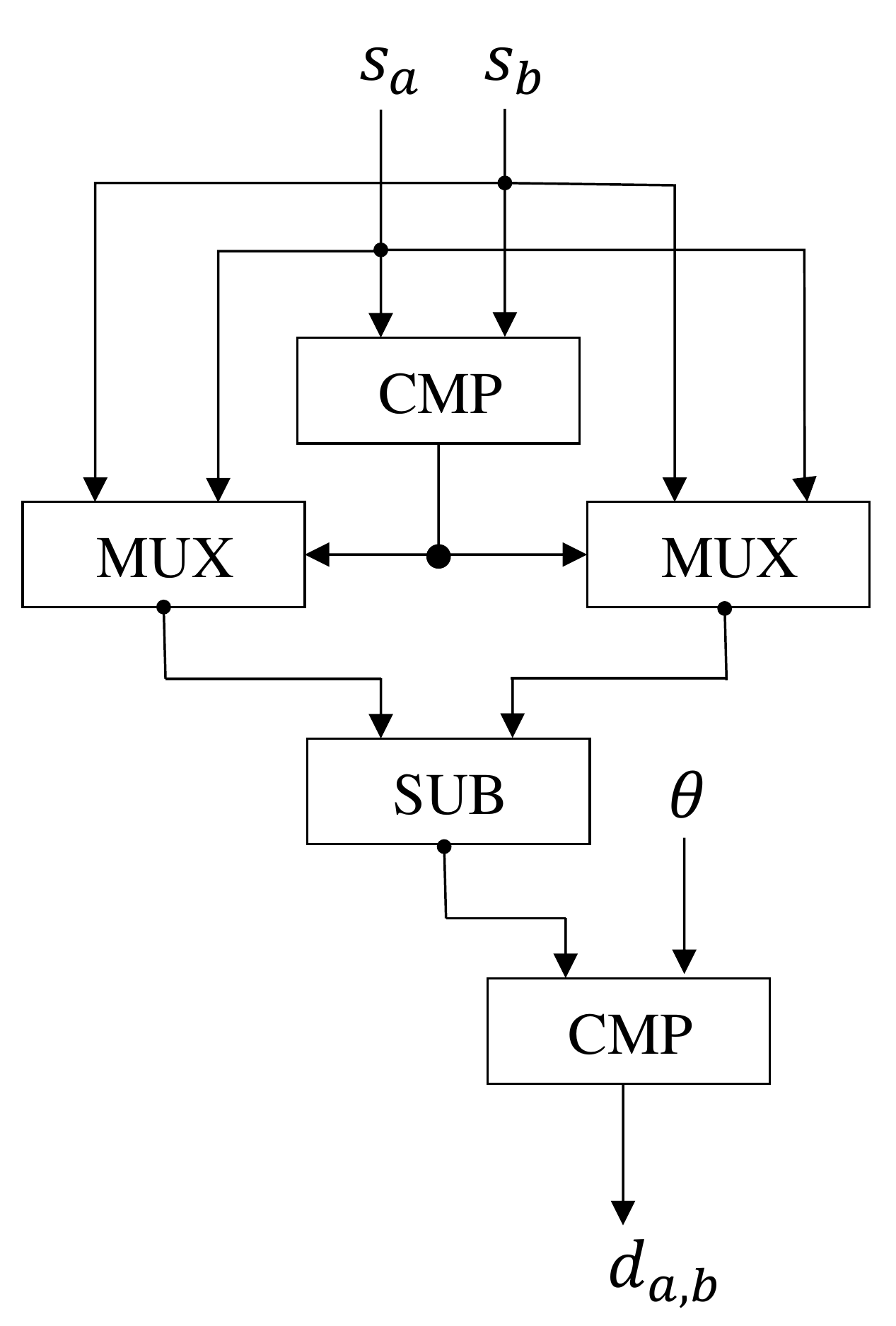}
\caption{Comparison circuit $\xi$.}
\label{circuit}
\end{figure}

At a high level, candidate $M_a$, the generator, first operates Algorithm \ref{al1} to encrypt its function $\xi$ into the garbled one $\hat{\xi}$ and then sends $\hat{\xi}$ together with the garbled input value $k_a$ of its input $s_a$ to other voters, who are deemed as the evaluators. To evaluate and vote on ${\it USM_a}$, evaluator $M_b$ runs Algorithm \ref{al2} which firstly operates the {\it 1-out-2} Oblivious Transfer protocol (OT) \cite{OT} to get the corresponding garbled input value $k_b$ of its private input $s_b$. Subsequently, $M_b$ performs $\hat{\xi}$ with $k_a$ and $k_b$ as the inputs, leading to the voting result. Basically, we design function $\xi$ as shown in Fig. \ref{circuit} \cite{guo}, which takes as inputs two user similarity values (i.e., $s_a$, $s_b$) and outputs the result whether the difference between them is less than $\theta$ or not (i.e., 1 or 0). Here, ${\it CMP}$, ${\it MUX}$, and ${\it SUB}$ respectively denote the comparator, multiplexer, and subtractor circuits. To be specific, the quantitative comparison result of $s_a$, $s_b$ can be obtained through ${\it CMP}$, which is then input into the ${\it MUX}$ circuits together with $s_a$, $s_b$ to produce the minuend and subtrahend for the following ${\it SUB}$ circuit. After that, by operating the ${\it SUB}$ circuit and comparing the subtraction result with $\theta$, we can derive the distance between two similarities as a result.

Algorithm \ref{Gen} describes the steps of generating a garbled gate, operated by the generator. Denote $w_1$ and $w_2$ as the inputs and $w_3$ as the output. For all possible values (i.e., 0 and 1) of $w_1, w_2$ and $w_3$, the generator first produces corresponding random values (Lines 1-6) leveraging the function $Random$. After that, the generator encrypts the random value of the output (i.e., $k_3^0$ and $k_3^1$) by utilizing the symmetric encrypt function $E$ with the random values of inputs (i.e., $k_1^0$, $k_1^1$, and $k_2^0$, $k_2^1$) as the secret keys (Lines 7-10). Through this, we can obtain the encrypted random values of the outputs $c_0^0$, $c_0^1$, and $c_1^0$, $c_1^1$, which are saved in a random order $g$ after being permutated by $\pi$, a random permutation function (Line 11). At last, the garbled gate $g$ is returned (Line 12) and we then finish the generating process. Notably, each gate in $\xi$ should be encrypted as described in Algorithm \ref{Gen} so as to finally reach the garbled circuit $\hat{\xi}$. In particular, $\hat{\xi}$ needs to be generated only once at the very beginning of consensus but the random value $k_a$ mapping to its input is required to be created in every round of voting. Again, since other miners only get the random value instead of the original data, the private data can be well-protected without any information leakage.

As for the circuit evaluator, it assesses the garbled circuit $\hat{\xi}$ to get the comparison result via running Algorithm \ref{Eva}. Specifically, the evaluator first gets the key $k_b$ mapped to its input value $s_b$ via employing the OT protocol (Line 1). Then, the garbled circuit $\hat{\xi}$ with $k_a$ and $k_b$ as the inputs is performed, which outputs the comparison result $\hat{d}_{a,b}$ (Line 2). Finally, the evaluator shares $\hat{d}_{a,b}$ with $M_a$ to get the comparison result $d_{a,b}$ through the mapping function (Line 3). This indicates the voting result of $M_b$ on $M_a$ is obtained (Line 4), based on which, the voter can determine the voting matrix accordingly.

\begin{algorithm}\label{al1}
{
    \caption{Garbled circuit generating algorithm $\mathsf{GEN}$}
    \label{Gen}
    \begin{algorithmic}[1]

        \REQUIRE~~\\
         $w_1,w_2$: circuit inputs\\
         $w_3$: circuit output

        \ENSURE~~\\
         $g=(c_0^0,c_0^1, c_1^0, c_1^1)$: garbled gate\\

       \STATE $k_1^0\gets Random(w_1=0)$
       \STATE $k_1^1\gets Random(w_1=1)$
       \STATE $k_2^0\gets Random(w_2=0)$
       \STATE $k_2^1\gets Random(w_2=1)$
       \STATE $k_3^0\gets Random(w_3=0)$
       \STATE $k_3^1\gets Random(w_3=1)$

       \STATE $c_0^0=E_{k_1^0}(E_{k_2^0}(k_3^0))$
       \STATE $c_0^1=E_{k_1^0}(E_{k_2^1}(k_3^0))$
       \STATE $c_1^0=E_{k_1^1}(E_{k_2^0}(k_3^1))$
       \STATE $c_1^1=E_{k_1^1}(E_{k_2^1}(k_3^1))$

       \STATE $g=\pi(c_0^0, c_0^1, c_1^0, c_1^1)$

       \STATE return $g$

    \end{algorithmic}}
\end{algorithm}

\begin{algorithm}\label{al2}
{
    \caption{Circuit evaluating algorithm $\mathsf{Eval}$}
    \label{Eva}
    \begin{algorithmic}[1]

        \REQUIRE~~\\
         $\hat{\xi}$: garbled circuit\\
         $k_a$: the key of input value $s_a$\\
        \ENSURE~~\\
         $d_{a,b}$: the comparison result of value $s_a$ and $s_b$\\

       \STATE Run 1-out-2 OT protocol to obtain $k_b$ for its input $s_b$
       \STATE $\hat{d}_{a,b}=\hat{\xi}(k_a, k_b)$
       \STATE Exchange $\hat{d}_{a,b}$ with the generator and get the comparison result $d_{a,b}$
       \STATE return $d_{a,b}$

    \end{algorithmic}}
\end{algorithm}

\subsection{Bayesian Truth Serum-Based Incentive Mechanism}\label{Sec-baysian}
The voting process may trigger liars, who grant their polls to the candidates that are beneficial to them, instead of the ones that are qualified. This egotistic behavior ruins the security of consensus because such a leader is incompetent to calculate similarity accurately. To suppress unfaithful voters, we propose an incentive mechanism based on Bayesian truth serum to elicit truth-telling behaviors \cite{bayesian}. The intuition behind our mechanism is that only through reporting the true beliefs of voting, can the miner gets the optimal revenue. By doing so, as rational miners pursuing personal payoff maximization, there is no motivation for them to hide their real beliefs.

Generally speaking, each miner $M_r$ $(r\in\{1,2,...,m\})$ is required to vote on $s_{i,j}$ of $M_i$, where $(i\in\{1,2,...,m\}, j\in\{1,2,...,n^2\})$, and at the same time, specify a prediction of the empirical distribution of the voting results. However, the true belief of $M_r$, i.e., $T^r\in\{0,1\}$, is unknown to others, so is the true distribution of the voter predicts, which is denoted as $\Omega_r$. Basically, we can get the posterior belief  $Pr(\Omega_r|T^r)$ from the common prior $Pr(\Omega_r)$ via $Pr(\Omega_r|T^r)\leftarrow Pr(\Omega)$ according to the Bayes formula and our voting mechanism is based on the assumption: the voter believes that other miners sharing the same vote would make the same inferences about the distribution accordance with itself, which can be described in the following.

\begin{assumption}
$Pr(\Omega_i|T^i)=Pr(\Omega_r|T^r)$ if and only if $T^i=T^r$ \cite{bayesian}.
\end{assumption}

For each $s_{i,j}$ submitted by $M_i$, miner $M_r$ is asked for two reports:
\begin{itemize}
 \item  \textbf{Voting report} $x_{ij}^{r} \in\{0,1\}$, which indicates $M_r$ supports the $j$-th similarity result of miner $M_i$ ($x_{ij}^{r}=1$) or not ($x_{ij}^{r}=0$).
 \item  \textbf{Prediction report} $y_{ij}^{r}\in[0,1]$, which is the prediction of distribution for the $j$-th similarity result of miner $M_i$ being approved.
\end{itemize}

Assume the number of voters in each round is $m-1$. That is, all other miners will participate in voting except the candidate itself\footnote{Note that the number of voters in each round can be adaptively adjusted, but here we fix it for simplicity.}. Hence, we can get the average vote for $s_{i,j}$ as $\bar{x}_{ij}=\frac{1}{m-1}\sum_{r=1}^{m-1}x_{ij}^{r}$. Accordingly, we set $s_{i',j}$ with $i'=arg \max \limits_{1\le i' \le m}\bar{x}_{i'j}$ as the entries of the global best user similarity. Besides, we define the number of votes $M_i$ as:
\begin{equation}
\label{eq2}
V_i=\sum_{j=1}^{n^2}\bar{x}_{ij}.
\end{equation}

To resist untruthful telling, we score each voter based on its voting and prediction reports. Given the possible real intention $\omega\in\{0,1\}$ and the prediction $y$ regarding the probability of $\omega=1$, a strictly proper scoring rule \cite{strict} defined as $H(y,\omega)$ can be expressed in the binary quadratic manner, which is,
\begin{equation}
\begin{split}
\label{score}
H(y,\omega=1)&=2y-y^2,\\
H(y,\omega=0)&=1-y^2.
\end{split}
\end{equation}
According to the robust Bayesian truth serum mechanism (RBTS) proposed in \cite{bayesian-2}, for miner $M^r$, it is required to select a {\it{reference}} miner $M^{\tilde{r}}$ with $\tilde{r}=r+1 (mod \ m)$ and a {\it{peer}} miner $M^{\vec{r}}$ with $\vec{r}=r+2 (mod \ m)$. Then, based on $y_{ij}^{r}$ reported by $M^r$, we can calculate
\begin{equation}
\label{pre-score}
y_{ij}^{r'}=\left\{
\begin{aligned}
y_{ij}^{\tilde{r}}+\delta,\ if\ x_{ij}^r=1,\\
y_{ij}^{\tilde{r}}-\delta,\ if\ x_{ij}^r=0,
\end{aligned}
\right.
\end{equation}
where $\delta=\min(y_{ij}^{\tilde{r}}, 1-y_{ij}^{\tilde{r}})$. In light of this, the reward that $M^r$ obtains from voting, i.e., $U^r$, can be described as
\begin{equation}
\label{utility}
U^r=\sum_{i=1,i\ne r}^{m}(\sum_{j=1}^{n^2}H(y_{ij}^{r'},x_{ij}^{\vec{r}})+\sum_{j=1}^{n^2}H(y_{ij}^{r},x_{ij}^{\vec{r}})).
\end{equation}
The first part of \eqref{utility} is called the information score, which is affected by the voting report $x_{ij}^r$ and the second part is termed as the prediction score \cite{bayesian-2}, changing with the prediction report $y_{ij}^r$. Hence, to investigate the reward maximization of \eqref{utility}, we proceed to analyze the best voting and prediction reporting mechanism separately, which finally indicates that the truth-telling strategy is the optimal one. Note that if the rewarding function is strictly proper, the best prediction mechanism is to report truthfully. Hence, in the following, we will only illustrate how our mechanism motivates miners to submit their voting reports truthfully. We abbreviate the voting choice submitted by a miner as $x$, the prediction as $y$, and let $p\in[0,1]$ as the true belief about the voting result, then the expected information score can be derived as
\begin{equation}\label{EP}
E[p]=p\times H(p,\omega=1)+(1-p)\times H(p,\omega=0).
\end{equation}
When the prediction is $y$, the miner's expected score can be obtained as
\begin{equation}\label{EY}
E[y]=p\times H(y,\omega=1)+(1-p)\times H(y,\omega=0).
\end{equation}
Hence, the expected loss of expected information score is
\begin{equation}\label{loss}
E[p]-E[y]=(p-y)^2,
\end{equation}
where we recognize $y$ as a parameter in the incentive mechanism. To minimize the expected loss, we should set $y$ appropriately so as to satisfy the absolute difference between $p$ and $y$, i.e., $|p-y|$, to reach the minimum. Now we testify that the scoring rule (\ref{pre-score}) can guarantee the above requirement, thus validating the effectiveness of our mechanism by corroborating that it is incentive compatible \cite{bayesian-2}, which is summarized in Theorem \ref{5.1}.

Suppose each miner forms a posterior belief $p\in\{p_{\{1\}},p_{\{0\}}\}$ about the prediction probability of voting as approval based on its real intention of 1 or 0.  And it always holds that $0<p_{\{0\}}<p_{\{1\}}<1$ for all admissible priors\cite{bayesian-2}. Then we have,
\begin{theorem}\label{5.1}
A miner can maximize its expected score by truthfully reporting the voting choice if $y\in(p_{\{0\}},p_{\{1\}})$.
\end{theorem}
\begin{IEEEproof}
We will prove that when the true voting report is 1, and the posterior is $p_{\{1\}}$, $y'=y+\delta$ can minimize the distance between it and $p_{\{1\}}$,  maximizing the expected score consequently. That is to say, $y+\delta$ has a shorter distance with $p_{\{1\}}$ compared to $y-\delta$ when the true voting report is 0. In this way, we have
\begin{itemize}
\item When $y+\delta \le p_{\{1\}}$, since $y>0$ and $\delta >0$, we can get $y-\delta < y+\delta \le p_{\{1\}}$.
\item Otherwise, if $y+\delta > p_{\{1\}}$, since $y<p_{\{1\}}$, we conclude that $(y+\delta)-p_{\{1\}}<p_{\{1\}}-(y-\delta)$.
\end{itemize}

The same procedure can be conducted when the true voting report is 0, thus we omit it for brevity.
\end{IEEEproof}


\subsection{Vote Counting Mechanism}
After incentivizing miners to truthfully report the votes of each candidate, PoUS steps to the vote counting phrase, which is committee-based. To be specific, PoUS first selects a vote-counting committee, denoted as $\mathcal{C}$. Particularly, the selection of committee\footnote{Note that the committee can be selected at the beginning of each round. In addition, to reduce the impact of committee alteration on PoUS, the members in committee can be adjusted after several rounds.} members can follow multiple rules according to the network size. If a large number of miners exist in the system, the random sampling functions are preferred, like follow the satoshi \cite{ran1}, verifiable random function (VRF) \cite{ran2}, since the random probability-based mechanisms can cut down communication costs while guaranteeing distribution. Otherwise, adopting the democratic voting schemes \cite{eos} is more favorable. Then, all the miners send their polls to $\mathcal{C}$ by running a smart contract illustrated in Algorithm \ref{a3}, regulated by the Voting and Voting-Result-Waiting timers. When the Voting-Result-Waiting timer expires, all the members in $\mathcal{C}$ start to count the votes and run a specific consensus mechanism locally, like Practical Byzantine Fault Tolerance (PBFT)\footnote{Other consensus mechanisms in permissioned blockchain are also feasible. And we omit the specific description of PBFT since it has been well-analyzed \cite{PBFT1,PBFT2}.}, to reach an agreement on {\it who} should be the leader in this round and {\it what} are the global best user similarities. At last, the committee is responsible to notify the leader and send the best-calculated user similarity results to it. This triggers the packaging stage of the leader subsequently, which will be demonstrated in the following section. Note that after the block is bailed, it is required to be sent to the committee for verification. If passes, the block is then broadcasted to all miners, or otherwise discarded.

\begin{algorithm}[t]
{
    \caption{Vote-sending smart contract}
    \label{a3}
    \begin{algorithmic}[1]

        \REQUIRE~~\\
         $\mathcal{C}$: the selected vote-counting committee;\\
         Voting timer and Voting-Result-Waiting timer: timers for miners to vote and send votes to $\mathcal{C}$.\\

       \IF {Voting timer is over}
        \IF{Voting-Result-waiting timer does not expire}
            \STATE{Send voting result to $\mathcal{C}$}
        \ENDIF
       \ENDIF
    \end{algorithmic}}
\end{algorithm}



\section{Clustering-Based Packing}\label{Sec-packing}
Essentially, the packing stage can be divided into two phases: 1) the leader is first required to {\it cluster} the transactions in its mempool based on the global best user similarity; and 2) {\it pack} transactions into a block according to their priorities. 
Note that there is no specific restriction on how to cluster, as long as it meets the requirements of the corresponding scenarios. 
We introduce the transaction priority rule as
\begin{equation}\label{priority}
P_{tx_i}=a\times D_{tx_i}+b\times Fee_{tx_i}+c\times Sim_{tx_i},
\end{equation}
where $D_{tx_i}, Fee_{tx_i}$ and $Sim_{tx_i}$ refer to the waiting time, submitted fee and similarity\footnote{In PoUS, $Sim_{tx_i}$ is measured by the inverse ratio of the transaction distance between $tx_i$ and the cluster center it belongs to.} of transaction $tx_i$ with $a,b,c$ as the scaling parameters. Our packing mechanism benefits from the following three aspects:
\begin{itemize}
  \item \textbf{Enabling cohort analysis}. We can find that a higher $Sim_{tx_i}$  leads to a higher probability that $tx_i$ can be bailed into a block. In doing so, blocks are packaged with transactions from similar users, facilitating PoUS to realize cohort analysis of users and portray them spatially.
  \item \textbf{Achieving fairness}. To make the transactions with lower similarity have a chance to be packaged, the metrics of {\it delay} and {\it fee} are added. Such a non-single-factor packaging priority makes our mechanism suitable for more transactions with different levels of user similarity.
  \item \textbf{Solving the cold start problem}. Although the new users have lower similarities due to fewer interactions, their transactions can also be packaged into a block in time by lifting fees or waiting for a longer time, fixing the cold start problem as a result. Furthermore, our mechanism trains new users to become patient or generous, which in turn enhances a sustainable system.
\end{itemize}

\begin{figure}
\includegraphics[scale=0.6]{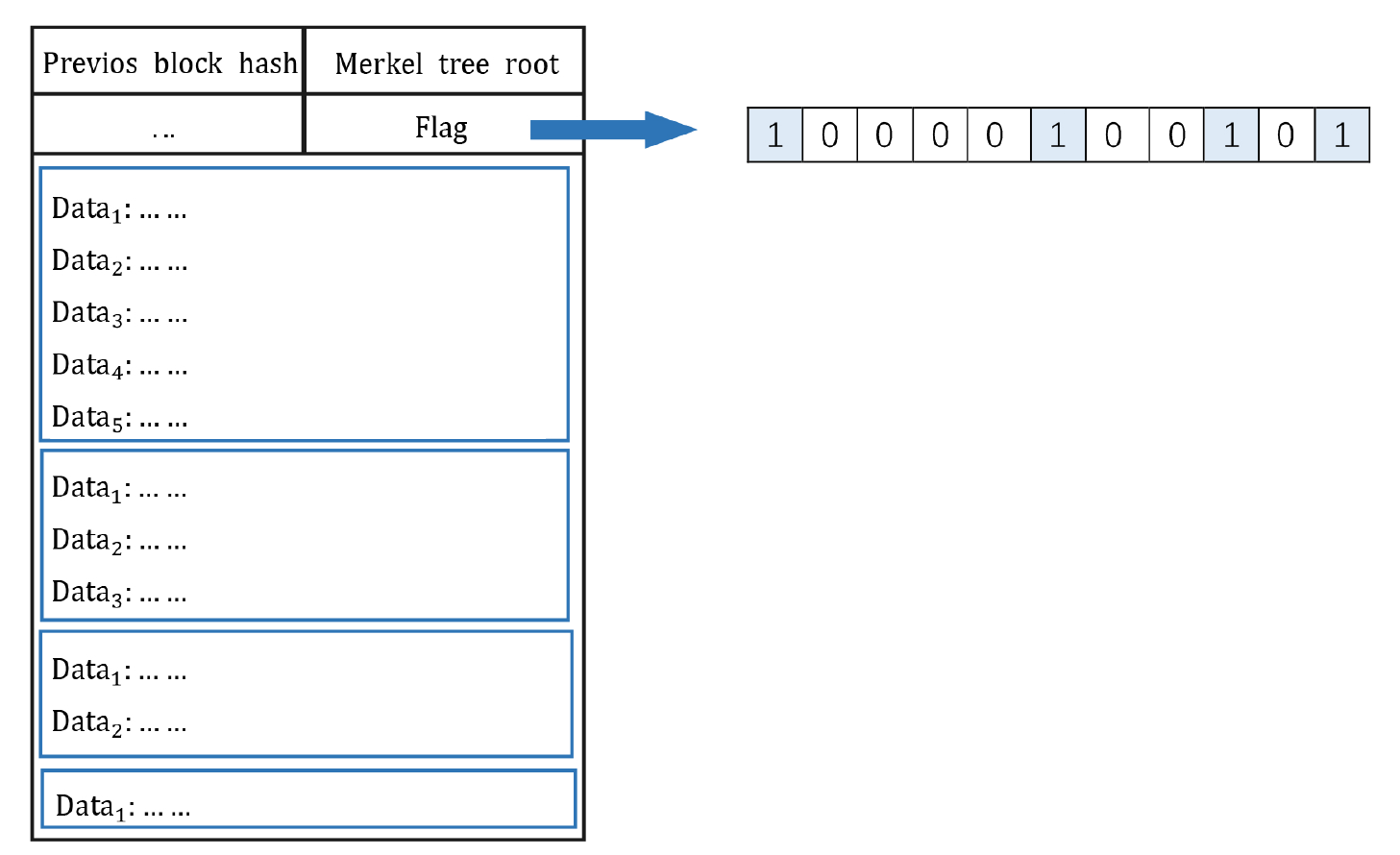}
\caption{The new designed block structure of PoUS.}
\label{blockheader}
\end{figure}

In order to represent the number of clusters a block includes, we add a {\it flag} field in the block header with keeping all other information in the current block structure, like the hash value of the previous block, the Merkle tree root of the included transactions, etc. The size of flag is $R$ bits, where $R$ denotes the maximum number of transactions each block can hold. The bit filled with 1 or 0 indicates whether its corresponding transaction stored in the block body is the first one of a cluster or not. Hence, we can say that the first bit of the flag is always 1, and it has at least one 1. In this way, we can count how many clusters the block has through visiting {\it flag} in the block header. An example of a block can be illustrated in Fig. \ref{blockheader}, where $R=11$ and there are four clusters with 5,3,2,1 transaction(s) included respectively. Hence, the flag can be then depicted as 10000100101.

\section{Evaluation}\label{Sec-eva}
In this section, we first present the security analysis for PoUS, which demonstrates its safety and liveness for being a robust and secure consensus mechanism. After that, we develop a prototype of PoUS in Python for experimental evaluation, to verify its performance, functionality and cost.

\subsection{Security Analysis}
For security concerns, we prove that PoUS satisfies the  {\it safety} and {\it liveness} properties, whose definitions are given as follows \cite{pyramid}:
\begin{definition}[Safety]
{\it PoUS can guarantee safety if the honest nodes agree on the same valid block in each round.}
\end{definition}

\begin{definition}[Liveness]
{\it PoUS can guarantee liveness if every block proposed by the leader in each round will eventually be committed or aborted.}
\end{definition}
\begin{theorem}
{\it The proposed PoUS can achieve safety and liveness when there are at least more than $\frac{2}{3}$ fraction of honest nodes in the vote-counting committee $\mathcal{C}$.}
\end{theorem}
\begin{IEEEproof}
Given more than $\frac{2}{3}$  honest nodes in $\mathcal{C}$, each member in the committee can reach consistency on {\it who is the leader}  and {\it whether the block proposed by the leader is legal or not} due to the adoption of PBFT consensus mechanism. Based on this, PoUS can realize both safety and liveness since other nodes not in the committee are only required to accept the result, i.e., the legal block.
\end{IEEEproof}

\subsection{Experimental Results}
\subsubsection{Implementation}
We build a prototype of PoUS in Python based on BlockSim, where the whole process including mining, voting, and packing is plugged. We compare the obtained results with those of PoW to show the superiority of PoUS. All the experiments are carried out on the machine with Intel Core i7-8700 GPU, 3.20 CPU and 8 RAM, and each simulation run is repeated 100 times to obtain the average value for statistical confidence.
\subsubsection{Setup}
We construct blockchain networks with $N=30, 200, 1000$ nodes\footnote{In the experiment part, we do not differentiate the identities of ``miner" and ``user", and uniformly represent them as ``nodes". That is, a node can serve the system as a miner, or benefit from the system as a user, or hold both characters by switching between different roles under different cases.}, with each mining power ranging in $[0,100]$ randomly. Besides, the transmission delay between each pair of nodes, the number of transactions each node produces, and the corresponding transaction fees all follow normal distribution parameterized with mean $\mu=0.4, 30, 0.000062$ and variance $\sigma=1$. 
Additionally, we categorize transactions into 6 classes, denoted as $\mathcal{A,B,C,D,E,F}$, respectively. Each class represents a kind of transaction, such as money transfer, smart contract creation, smart contract invocation, etc\footnote{ Different blockchain-based systems may possess different classification standards, which will not be discussed in this paper.}. In the stage of mining, each node first collects transactions from its mempool and the latest block (i.e., $\eta=1$), and then counts the number of transactions for each type generated by each user $U_i,i \in \{1,...N\}$, which are represented as $N_\mathcal{A}^i$ to $N_\mathcal{F}^i$. Accordingly, we can assemble them as the user vector $F^i=(N_\mathcal{A}^i,N_\mathcal{B}^i,...,N_\mathcal{F}^i)$  for calculating user similarity by exerting Euclidean distance with the recalculation update scheme. As for the voting stage, the garbled function is realized as described in Section \ref{Section-encryption}. We employ the $\mathsf{SHA256}$ hash function to facilitate symmetric encryption with keys distributed at the beginning of system operation. Each user will go through the 2PC-based voting process and be incentivized via the Bayesian truth serum-based mechanism. When the similarity gap is less than $\theta= 0.4$, a vote labeled with ``1" will be cast, or otherwise, it will be labeled with ``0".

\subsubsection{Performance evaluation}
We measure the performance of PoUS via two metrics: 1) transaction per second (TPS) and 2) transaction confirmation latency, with various block sizes and block intervals. After obtaining the results, we compare them with those of PoW. The environmental configuration and experimental parameter settings are shown respectively in Tables \ref{table-ex1} and \ref{table-ex2}.

\begin{table}[t]
	\centering
	\caption{Environmental configuration settings.}
    \label{table-ex1}
	\begin{tabular}{cccc}
		\toprule  
		Parameter&Value&Unit&Description \\
		\midrule  
		Sim Time&10000&s&Simulation running time \\
        \midrule  
		Transaction Size&250&Byte&Average size of a transaction\\
        \midrule  
		Transaction Delay &0.5&s& Delay for transmitting a transaction\\
        \midrule  
        Breward&6.25&BTC&Block reward\\
		\bottomrule  
	\end{tabular}
\end{table}

\begin{table}[t]
	\centering
	\caption{Experimental Parameter Settings.}
    \label{table-ex2}
    \begin{threeparttable}
	\begin{tabular}{cccc}
		\toprule  
		Metric&Value\tnote{1}&Unit&Description \\
		\midrule  
		Network Size&30$\sim$1000&$/$&Number of nodes\\
        \midrule  
		Block Size&0.5$\sim$16&MB&Size of a block\\
        \midrule  
		Block Interval &200$\sim$1000&s&Block generation time gap\\
        \midrule  
        Block Delay&0.2$\sim$0.6&s&Delay for transmitting a block\\
		\bottomrule  
	\end{tabular}
   \begin{tablenotes}    
        \footnotesize               
        \item[1] The metrics are sampled in this range.
      \end{tablenotes}            
    \end{threeparttable}       
\end{table}

\begin{figure*}
	\centering
	\subfigure[$N=30$]{
		\begin{minipage}[t]{0.3\linewidth}
			\centering
			\includegraphics[width=2in]{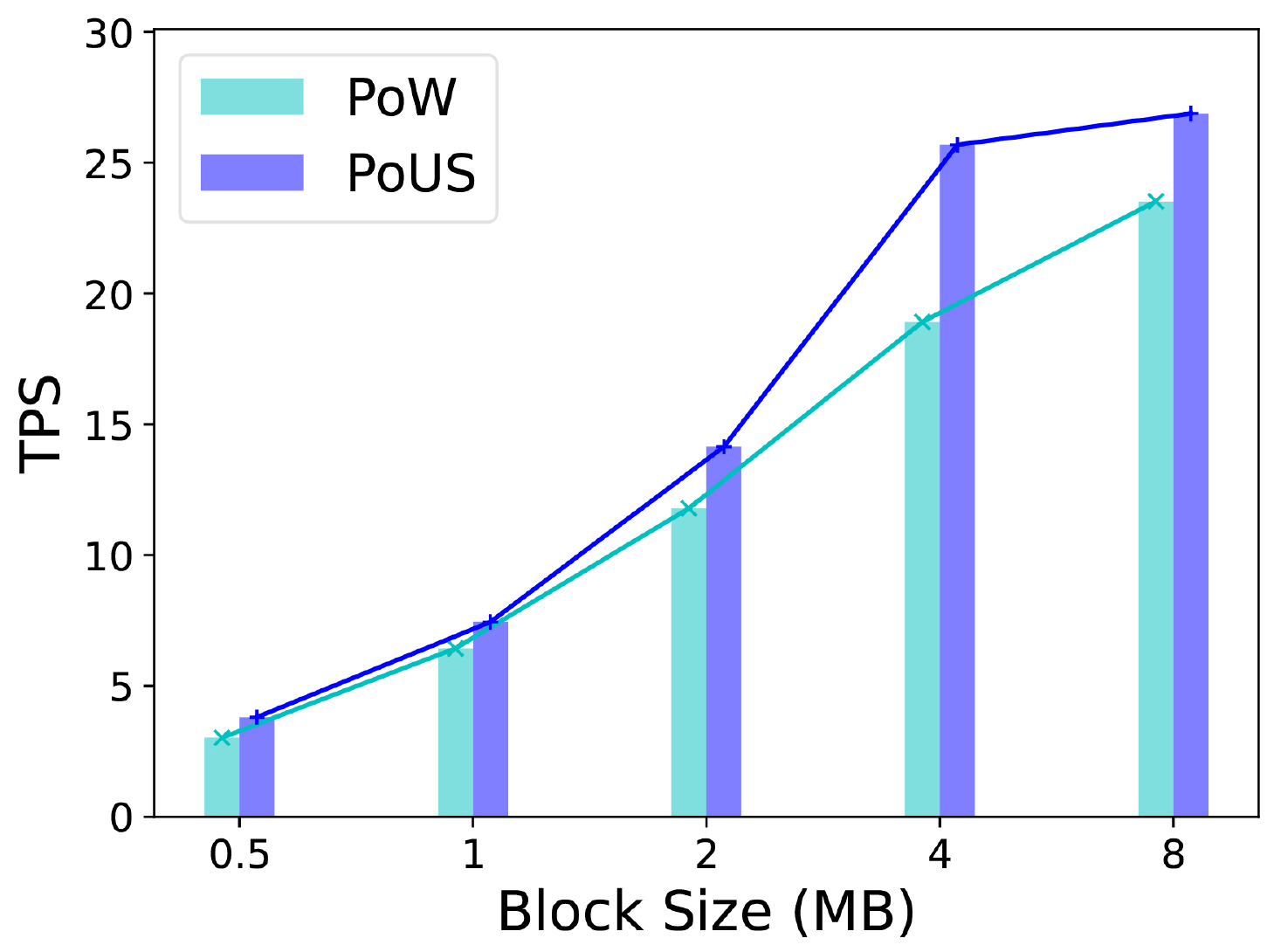}
		\end{minipage}
	}%
	\subfigure[$N=200$]{
		\begin{minipage}[t]{0.3\linewidth}
			\centering
			\includegraphics[width=2in]{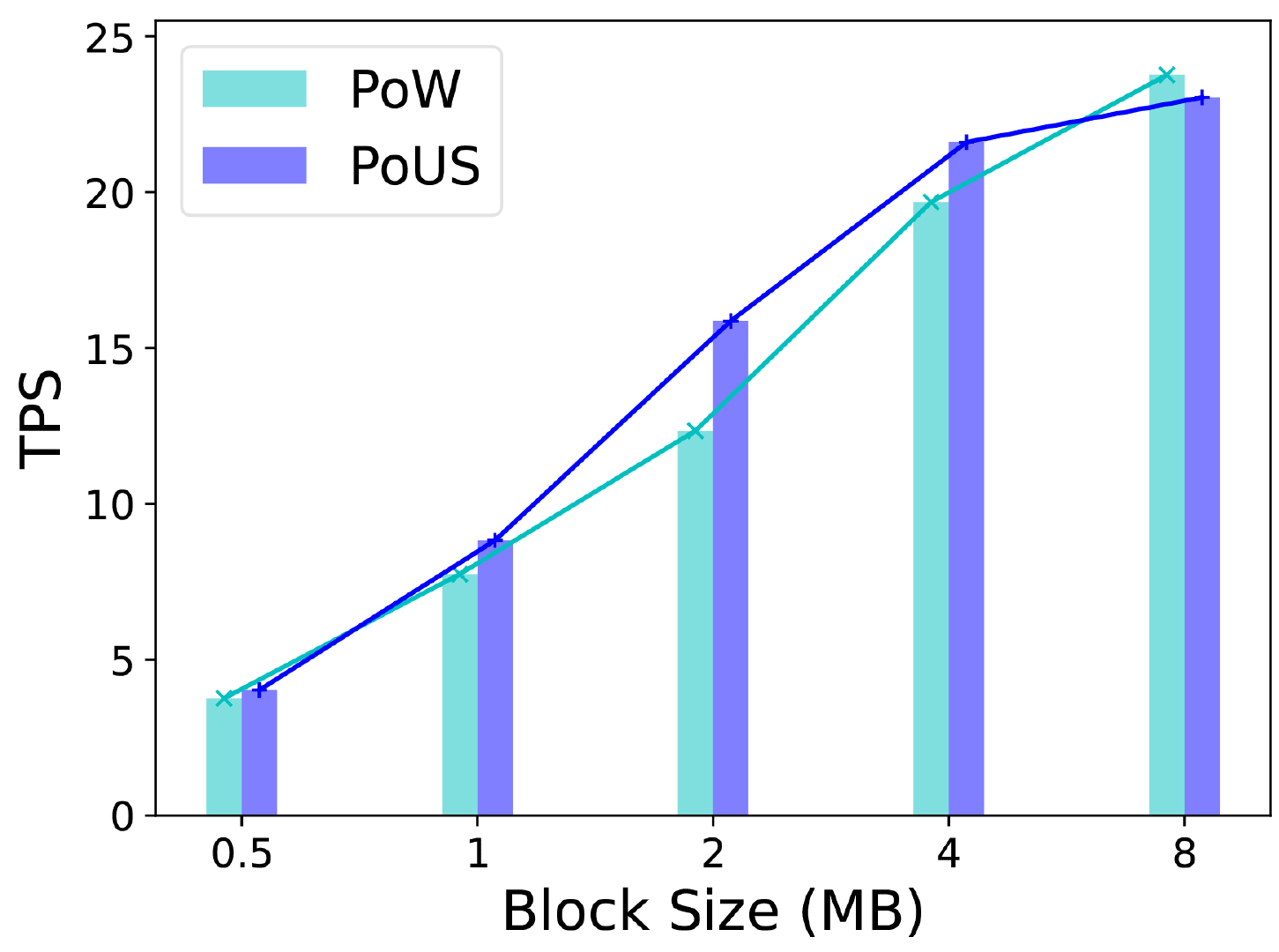}
		\end{minipage}
	}%
	\subfigure[$N=1000$]{
		\begin{minipage}[t]{0.3\linewidth}
			\centering
			\includegraphics[width=2in]{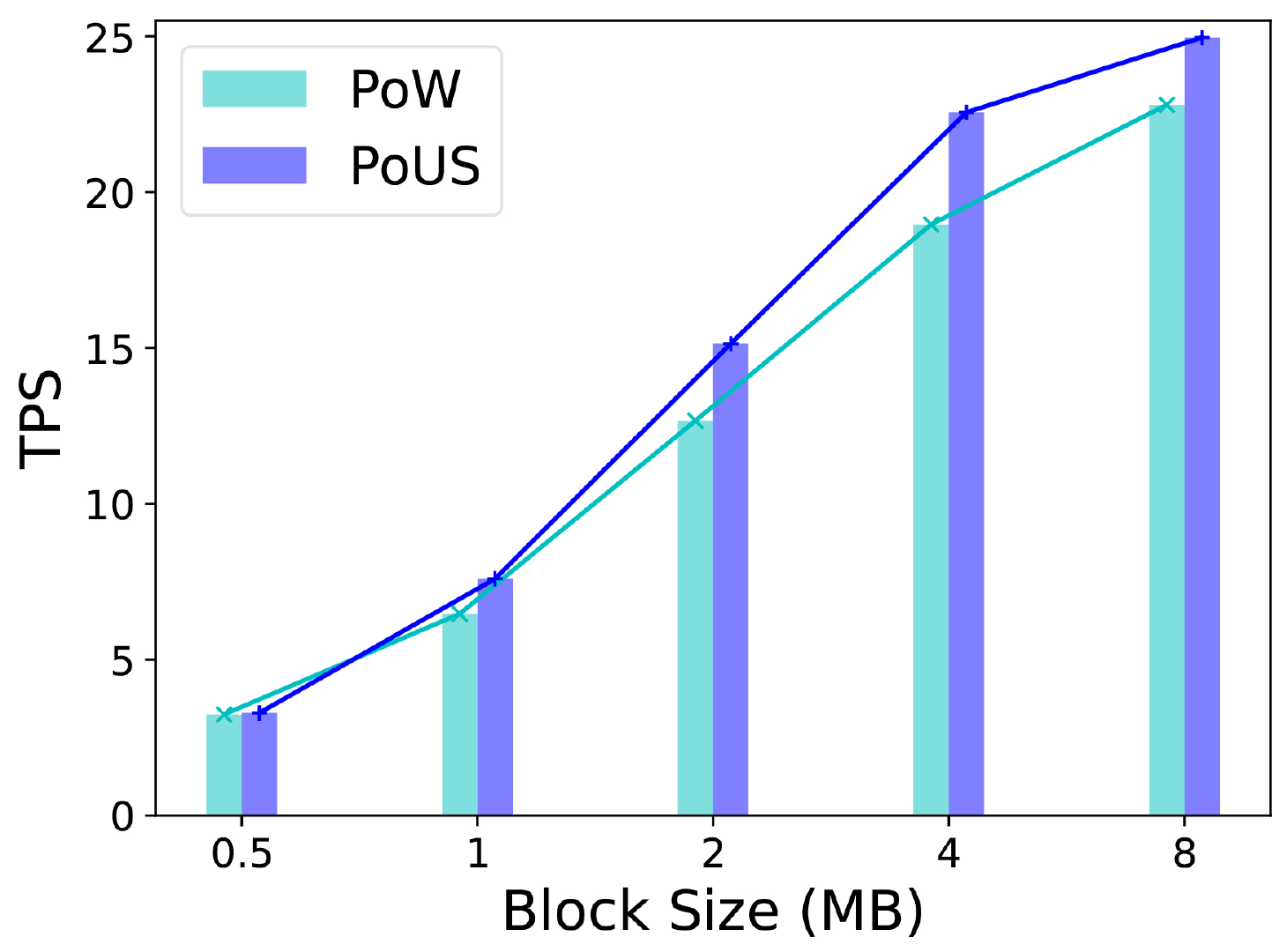}
		\end{minipage}
	}%

	\caption{Transaction per second (TPS) when varying block size with $N=30, 200, 1000$.}
	\label{fig:block-size-to-TPS}
\end{figure*}

Fig. \ref{fig:block-size-to-TPS} reports the number of confirmed transactions per second (TPS) for PoW and PoUS with varying block size when the block delay and block interval are set as 0.4s and 600s. Subfigures (a) (b), and (c) are conducted under different network sizes $N=30,200,1000$, respectively. From them, we can conclude that: 1) when the block size is 1MB, the TPS of PoW is nearly 7, which is exactly in line with the Bitcoin system in practice. This demonstrates that simulating PoW with BlockSim is credible, making it plausible and convincing to evaluate PoUS through BlockSim since they share the same environmental and experimental settings; 2) both the TPS of PoW and PoUS will increase as the rise of block size. Since the larger the block is, the more transactions it can hold, leading to the positive relationship between block size and TPS; 3) PoUS achieves an average TPS improvement of 15.66\% compared with PoW.

\begin{figure*}
	\centering
	\subfigure[$N=30$]{
		\begin{minipage}[t]{0.3\linewidth}
			\centering
			\includegraphics[width=2in]{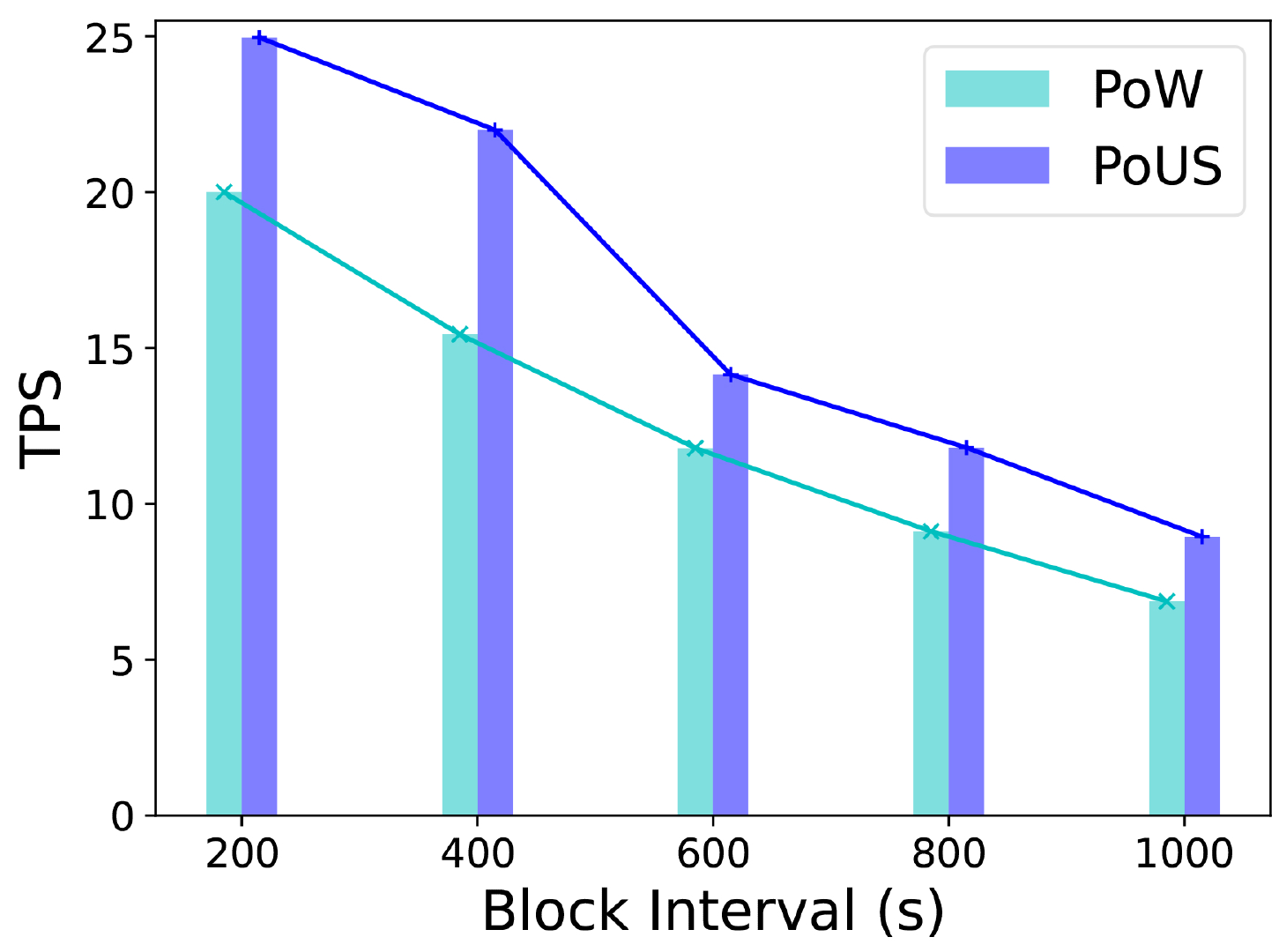}
		\end{minipage}
	}%
	\subfigure[$N=200$]{
		\begin{minipage}[t]{0.3\linewidth}
			\centering
			\includegraphics[width=2in]{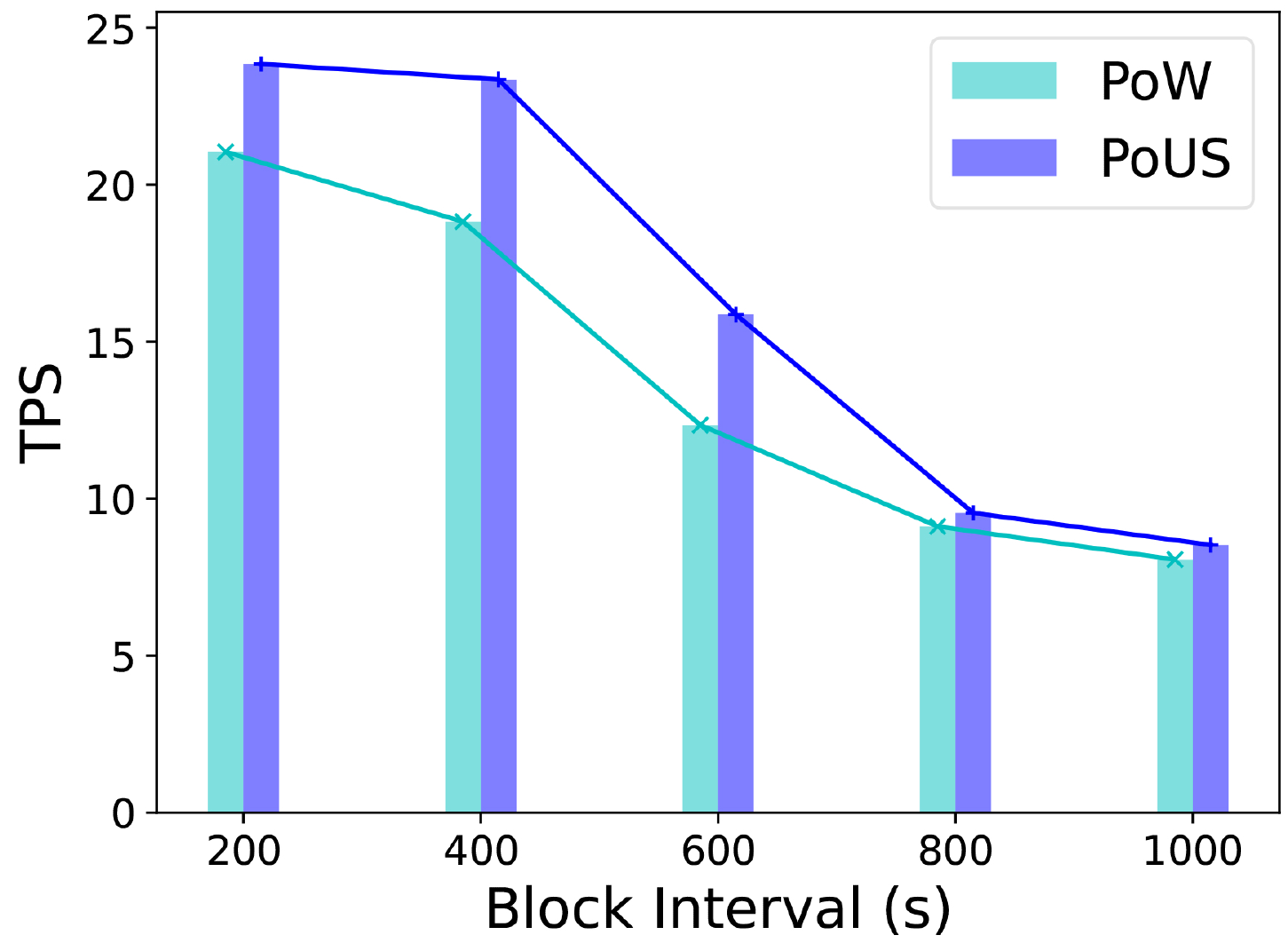}
		\end{minipage}
	}%
	\subfigure[$N=1000$]{
		\begin{minipage}[t]{0.3\linewidth}
			\centering
			\includegraphics[width=2in]{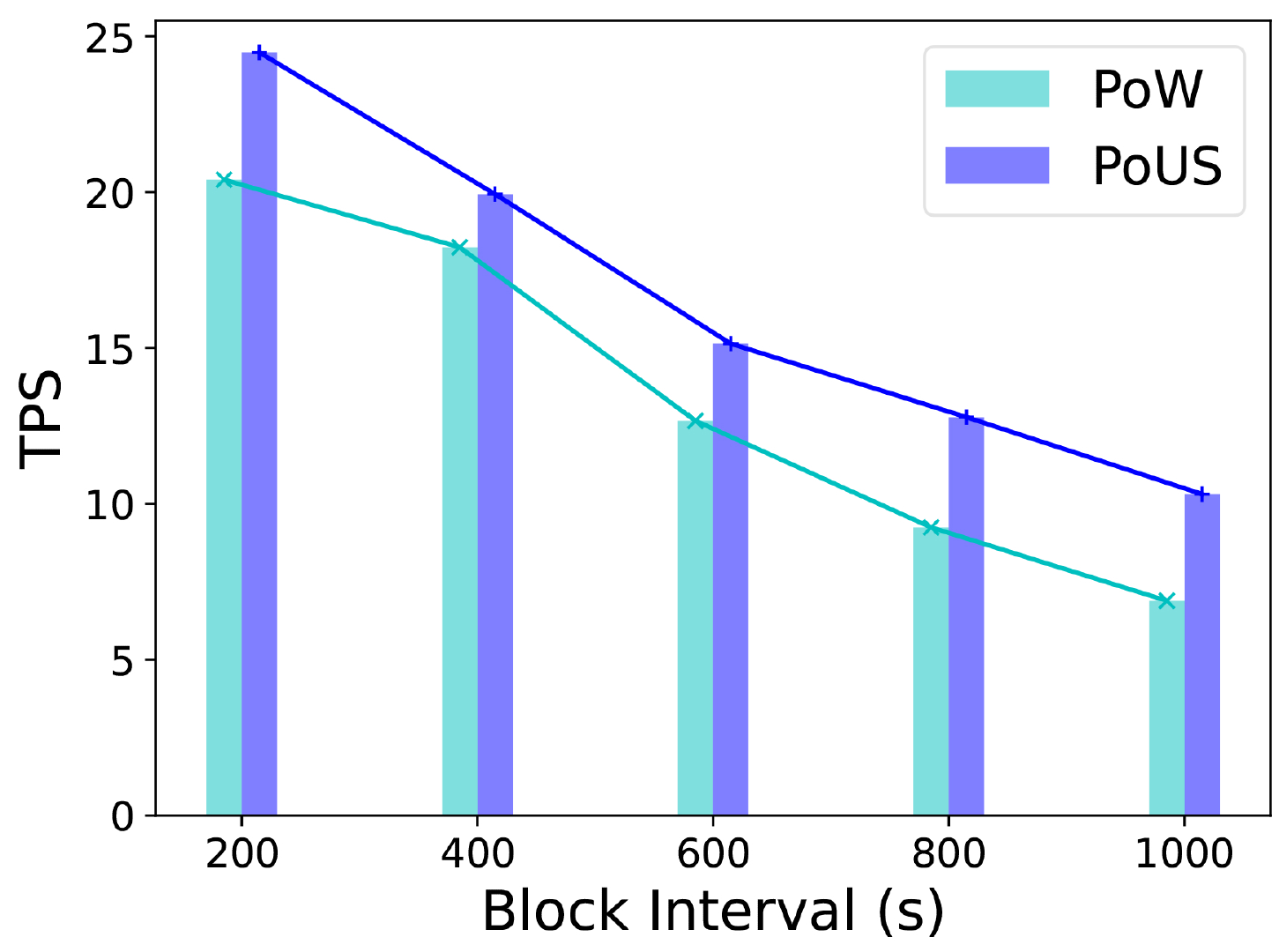}
		\end{minipage}
	}%

	\caption{Transaction per second (TPS) when varying block interval with $N=30, 200, 1000$.}
	\label{fig:block-interval-to-TPS}
\end{figure*}

Fig. \ref{fig:block-interval-to-TPS} plots the TPS of PoW and PoUS on the difference of block interval, as the block size and delay are set as 2MB and 0.4s. We present the results under different network sizes $N=30, 200, 1000$ in subfigures (a), (b), and (c), based on which, we find that: 1) block interval exerts a negative effect on TPS for both PoW and PoUS. This is straightforward since the longer the time gap between successful blocks, the fewer blocks will be put on blockchain per unit time, narrowing down the transaction throughput as a result; 2) PoUS surpasses PoW in transaction throughput by 24.01\% on average.

\begin{figure}
	\centering
	\subfigure[Confirmation latency VS. block size.]{
		\begin{minipage}[t]{0.8\linewidth}
			\centering
\includegraphics[width=2.5in]{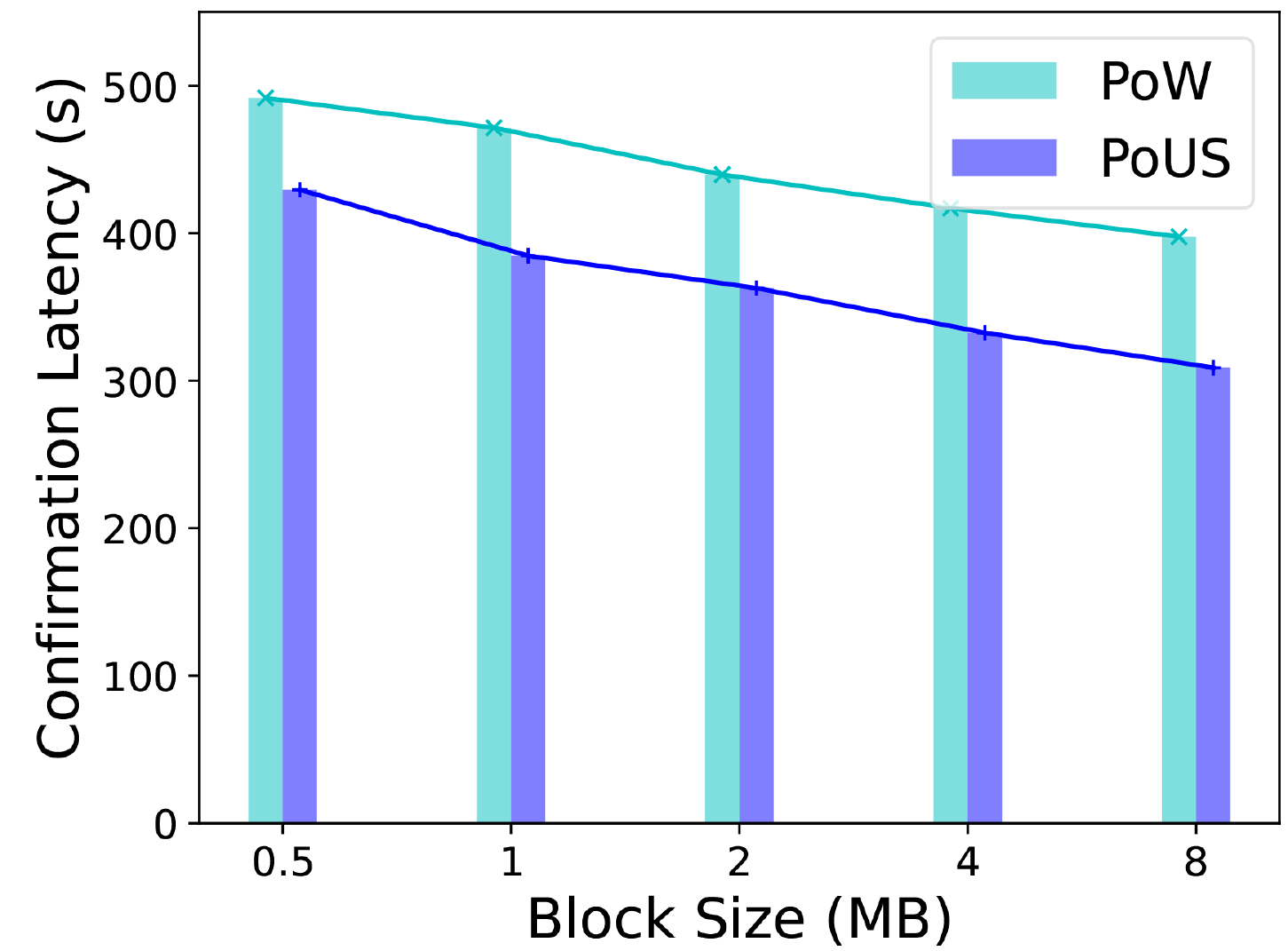}
		\end{minipage}
	}%

	\subfigure[Confirmation latency VS. block interval.]{
		\begin{minipage}[t]{0.8\linewidth}
			\centering \includegraphics[width=2.5in]{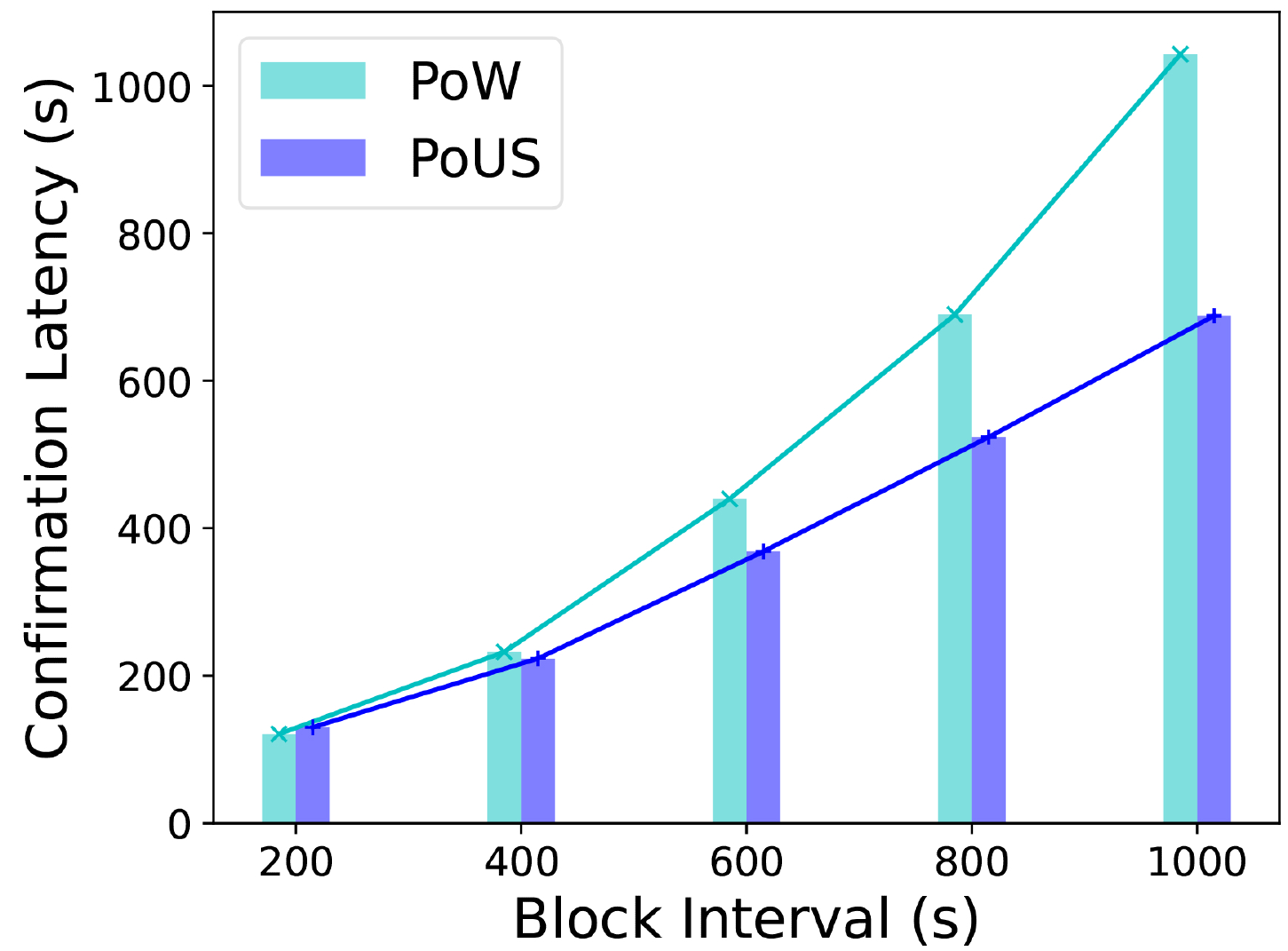}
		\end{minipage}
	}%
	\caption{The confirmation latency when varying block size and interval with $N=30$.}
	\label{fig:latency}
\end{figure}

We also evaluate the transaction confirmation latency in PoUS and PoW, followed by the predefined transaction priority \eqref{priority} where $a=0.5, b=2, c=1$. The comparisons are conducted under network size $N=30$\footnote{Note that extensive simulations with different network sizes are carried out, whose results demonstrate very similar trends, thus we omit to show them to void redundancy.}. Fig. \ref{fig:latency} shows the relationships between confirmation latency with different block sizes and intervals. From subfigure (a), we can point out that: 1) the confirmation latency dwindles as the increase of block size since a larger block can intake more transactions, reducing the latency as a consequence; 2) when the block size varies, PoUS reduces the latency by an average of 24.14\% compared with PoW. From subfigure (b), conclusions can be drawn that: 1) the latency increases with the growth of block interval. This is because larger intervals may shorten the number of confirmed transactions on the chain per unit time, thus bringing about a higher latency; 2) when the block interval changes, the latency of PoUS is less than that of PoW about 43.64\% on average. Furthermore, PoUS is interval-insensitive since the growth rate of confirmation latency with interval is linear while that of PoW shows an approximately exponential trend.

To sum up, we conclude the performance of PoUS by presenting the following observations:

{\it Observation 1: PoUS outperforms PoW in achieving an average TPS improvement of 15.66\% and 24.01\% when varying block size and interval.}

{\it Observation 2: PoUS surpasses PoW in dwindling the confirmation latency by an average of 24.14\% and 43.64\% when changing block size and interval. Moreover, PoUS is interval-insensitive.}
\subsubsection{Functionality evaluation}
We corroborate that PoUS can build up a detailed profile of users by packaging similar transactions into a block, so as to empower the consensus mechanism in measuring the spatial information of blockchain. To that aim, we need to compare two transaction sets: the first set represents transactions in clusters, which are determined by the global best user similarity. This set can be regarded as the baseline showing similar transactions; the second set contains transactions that the leader actually packages, that is, the selected transactions to be assembled currently. If these two sets are consistent, it means that the leader does package transactions with higher similarity, demonstrating PoUS functions as we wish.

\begin{figure}
\centering
\includegraphics[scale=0.45]{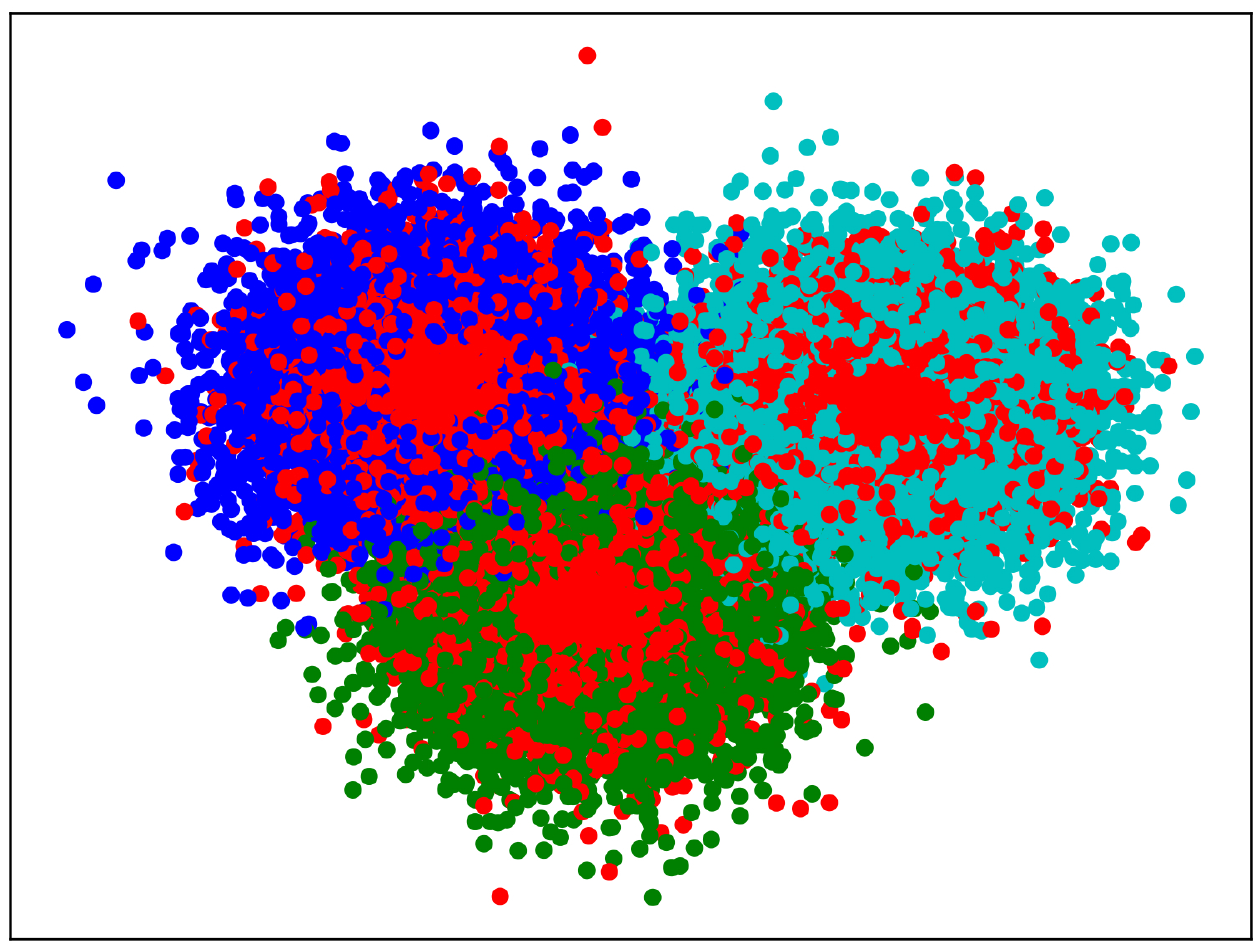}
\caption{The clustering results (i.e., the dots colored in blue, cyan, and green) VS. the selected transactions to be packed (i.e., the red dots).}
\label{clustering}
\end{figure}

We randomly select one leader as the ``target leader" during the simulation process for experimental purposes\footnote{Note that situations where other leaders are chosen demonstrate similar results, hence we omit them for brevity.}, and obtain the clustering result based on the source user of transactions in light of the k-means scheme. To illustrate more explicitly, we employ the Principal Component Analysis (PCA) to project the multidimensional vectors into two dimensions. The results are presented in Fig. \ref{clustering}, where the clusters are colored in blue, cyan, and green, and the selected transactions are painted in red. We can conclude that: 1) the red dots are basically concentrated in the middle of each cluster, which reflects that the transactions selected by PoUS are indeed of high user similarity. Hence, PoUS can reveal the spatial information of users in blockchain as we expected; 2) some red dots are around the periphery of each cluster. This suggests that even though some transactions are disadvantaged in similarity, they can also be selected if they are urgent or with high fees, reflecting the fairness of our mechanism.

Accordingly, we summarize the functionality of PoUS by presenting the following observation:

{\it Observation 3: PoUS can function well to portray the spatial information of users as we expected, and the transaction priority rule is fair.}

\subsubsection{Computation time and communication cost}
\begin{figure}
	\centering
	\subfigure[Computation time VS. data size.]{
		\begin{minipage}[t]{0.8\linewidth}
			\centering
\includegraphics[width=2.5in]{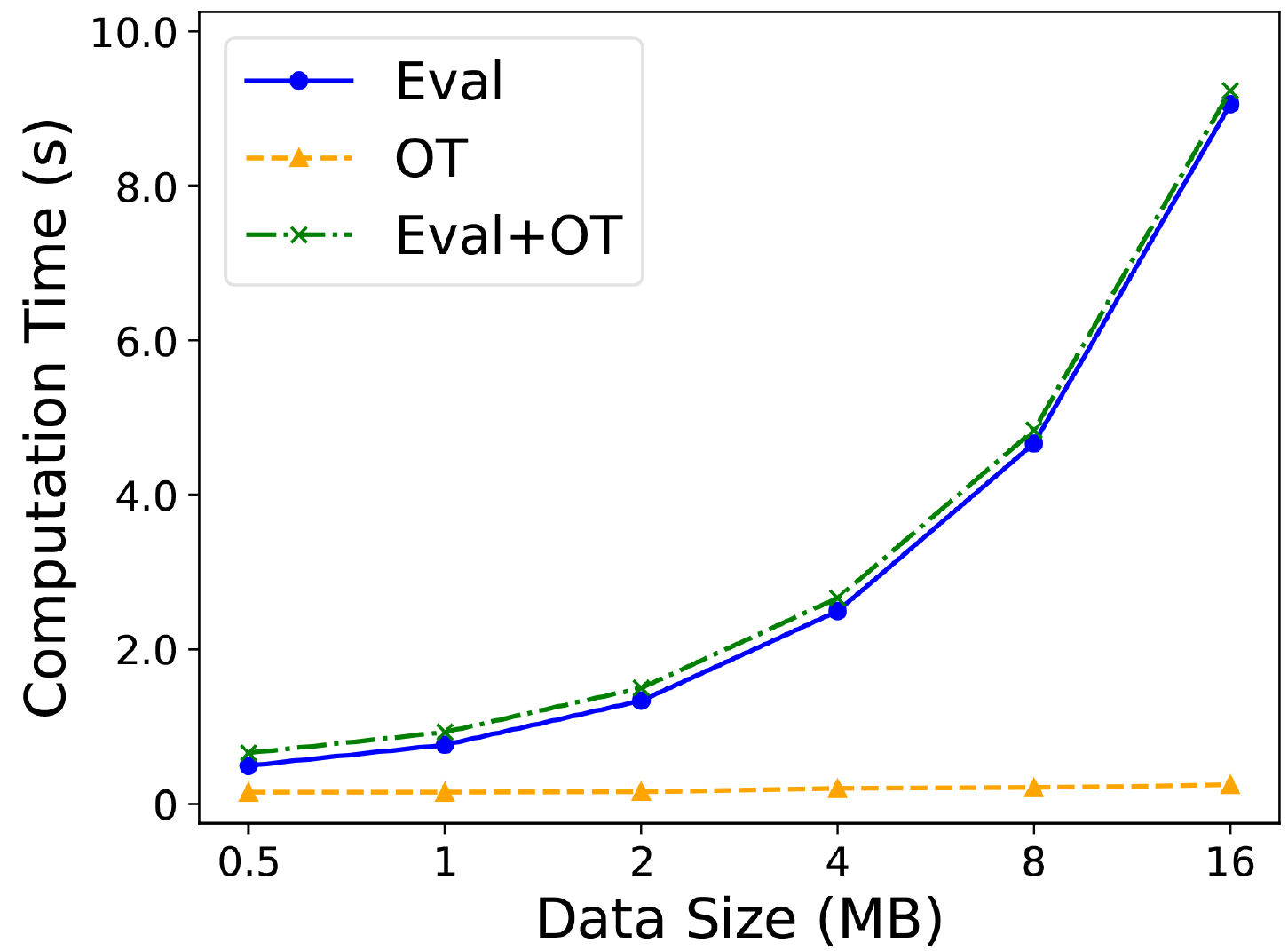}
		\end{minipage}
	}%

	\subfigure[Communication cost VS. the number of users.]{
		\begin{minipage}[t]{0.8\linewidth}
			\centering \includegraphics[width=2.5in]{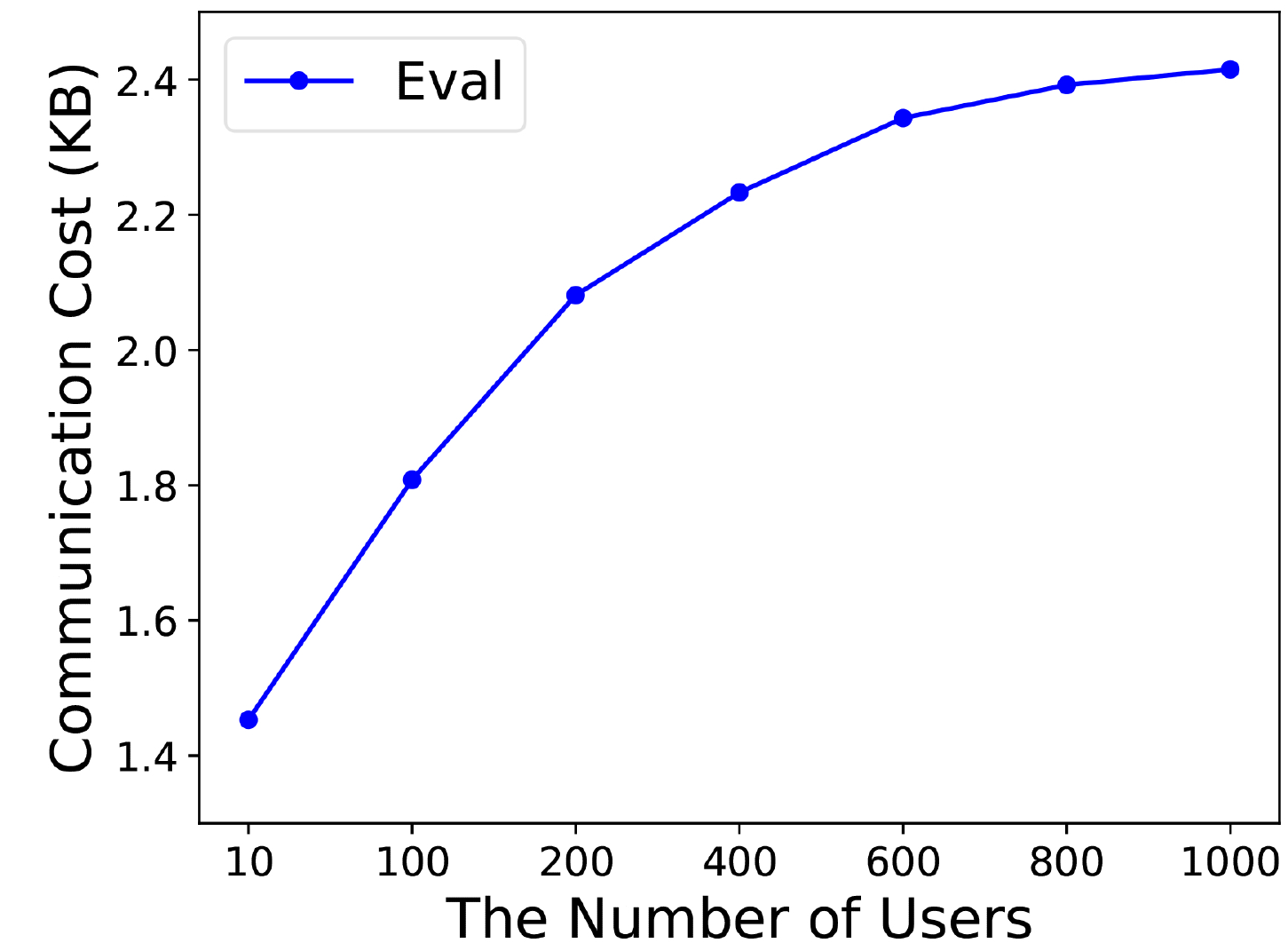}
		\end{minipage}
	}%
	\caption{The cryptographic computation time and communication cost.}
	\label{fig:cost}
\end{figure}

In this part, we evaluate the cryptographic computation time and communication cost during the 2PC-based voting process, as depicted in Fig. \ref{fig:cost}. In detail, Fig. \ref{fig:cost} (a) describes the exhausted computation time changing with the data size, which covers the OT execution  and circuit evaluation (Eval). From this, we find: 1) the time consumption mainly comes from Eval process and the data size lays a positive effect on it; 2) the computation time is negligible since the encryption time is less than 2s when the data size is 2MB. In addition, we plot the communication cost of Eval on the difference of user number, where the compressed row storage scheme is adopted. As described in Fig. \ref{fig:cost} (b), when the number of users in blockchain goes more, no more than 2.5 KByte communication cost will be spent, which is completely acceptable.

{\it Observation 4: The computation time and communication cost that PoUS consume are negligible, making it achievable in practice.}

\section{Other Concerns}\label{sec-concerns}
In this section, we list some concerns that readers may arouse about PoUS and present explanations, to show the opportunity for the wide application of PoUS in practice.

\textbf {Question 1:} {\it PoUS requires a smart contract, how to solve the chicken-egg problem where the smart contract is supposed to be secured by the consensus primarily?}

\textbf {Answer:} At the beginning of PoUS, we can exert other consensus mechanisms for permissionless blockchain, such as PoW, PoS, etc, to reach an agreement on the smart contract firstly. After the smart contract has been acknowledged by all the nodes, we then transfer to PoUS consensus.

\textbf {Question 2:} {\it How can PoUS defend against double-spending attack or other computation-based attacks?}

\textbf {Answer:} It is worth noting that PoUS is immune to the computation-based attacks since the leaders are selected in advance, followed by packaging similar transactions into a block. Hence, PoUS can reach deterministic finality rather than probabilistic one like PoW, making no room for the dominant nodes to overtake the main chain.

\textbf {Question 3:} {\it What are the advantages of establishing a searchable blockchain based on user similarity?}

\textbf {Answer:} Essentially, PoUS can achieve user/transaction classification in the consensus stage, which brings in the following two merits.
Firstly, PoUS facilitates a searchable blockchain from the transaction level. Through combining efficient storage and index schemes, it can support diverse querying requirements and greatly enhance the query efficiency. In doing so, accurate decision-making can be realized. To illustrate, for the cold-chain transportation transactions under the COVID-19 pandemic, if a transaction is confirmed to be epidemic-related, other possible risky transactions can be quickly located based on the clustering results of the source and end users (in this case, user similarity is defined as the frequency of transactions), so as to curb the spread of the disease. Secondly, the similarity-in-design consensus can also record the users' behaviors behind the generated transactions, which in turn facilitates cohort analysis of users. Cohort analysis can reveal the characteristics of users through the fancy data, analyzing the differences among various user groups, according to which, the key factors that affect user retention intentions can be found. In this way, we can quantify user value retention and uncover the effect of system improvement more comprehensively.

\section{Related Work}\label{Sec-rel}
Recently, many attempts have been devoted to developing greener consensus mechanisms as substitutes for PoW. Basically, there are two lines of improvements from the perspectives of {\it energy-conservation} and  {\it energy-recycling},
where our PoUS belongs to the latter. Hence we focus on investigating the energy-recycling studies as follows.

Initially, the meaningless nonce in PoW is replaced with some mathematic problems, such as prime number \cite{primecoin}, matrix-based issues \cite{PoX}, etc. Subsequently, more complicated problems designed from reality are presented. Zhang {\it et al.} \cite{rem} put forward a resource-efficient mining framework for blockchain, called REM, that utilizes the trusted hardware, i.e., the Intel Software Guard Extensions (SGX), to reinvest the wasted computations for executing useful downloads outsourced by clients. This promotes the idea of Proof of Useful Work (PoUW). However, the heavy reliance of functioning REM on SGX may violate the decentralization nature of blockchain, as stated in \cite{CN}. Hence, Lasla {\it et al.} in \cite{CN} proposed to divide time into epochs, with each comprising two consecutive mining rounds. And only the selected runner-ups can join in the second round to be compensated with block reward. In doing so, the number of competing blocks can be greatly reduced, which in turn narrows down the consumed computing energy. In parallel, Du {\it et al.} \cite{du} presented a novel mechanism that exploits PoW mining power to accelerate decentralized machine learning through scheduling tasks among multi-access edge computing servers. Additionally, Qu {\it et al.} in \cite{PoFL} repurposed the computing resource to federated learning, which has a natural fit in terms of the organization structure of pooled mining in blockchain. By doing so, they introduced Proof of Federated Learning (PoFL), together with a reverse game-based data trading mechanism and a privacy-preserving model verification mechanism enhanced by homomorphic encryption and 2PC techniques. Besides, Li {\it et al.} \cite{li} exploited the computation power of miners for biomedical image segmentation, based on which, a segmentation model training that can handle multiple tasks, larger models and training datasets was designed.


\section{Conclusion}\label{Sec-conclusion}
A novel energy-recycling consensus mechanism named {\it proof of user similarity} (PoUS) is proposed in this paper, where the valuable computing resource is reinvested to calculate the similarity of users. PoUS is designed with three stages, which are mining, voting and packing. Each of them respectively serves for similarity calculation, leader selection, and packaging blocks. To address the supply-demand contradiction, we embrace the best-effort schema to allow the miners to compute user similarities partially. Besides, considering the plagiarism and lying risks rooted in the voting process, we present a 2PC-based voting mechanism and a Bayesian truth serum-based incentive mechanism. The former can leverage the cryptographic primitives to assure correct voting without disclosing any private information about the candidates, while the latter encourages the profit-driven miners to honestly report their true beliefs. As for the packing period, we design a fair and effective transaction priority rule for selection.
We testify PoUS by implementing a prototype, whose results demonstrate that PoUS surpasses PoW in achieving an average TPS improvement of 24.01\% and an average confirmation latency reduction of 43.64\%. Besides, PoUS functions well with negligible computation time and communication cost in mirroring the spatial information of users, which can replenish blockchain besides the temporal scale from the spatial dimension.

\ifCLASSOPTIONcompsoc
  \section*{Acknowledgments}
\else
  \section*{Acknowledgments}
\fi
This work has been supported by National Key R$\&$D Program of China (No. 2019YFB2102600), National Natural Science Foundation of China (No. 62072044), the International Joint Research Project of Faculty of Education, Beijing Normal University, and Engineering Research Center of Intelligent Technology and Educational Application, Ministry of Education.

\bibliographystyle{IEEEtran}
\bibliography{reference}

\begin{IEEEbiography}[{\includegraphics[width=1in,height=1.25in,clip,keepaspectratio]{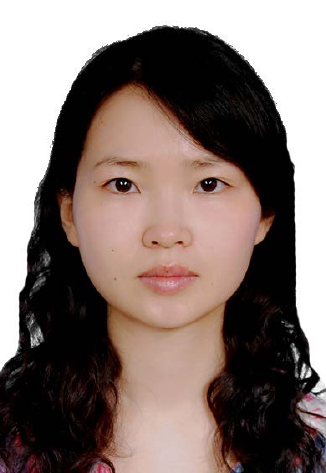}}]{Shengling Wang} is a full professor in the School of Artificial Intelligence, Beijing Normal University. She received her Ph.D. in 2008 from Xi'an Jiaotong University. After that, she did her postdoctoral research in the Department of Computer Science and Technology, Tsinghua University. Then she worked as an assistant and associate professor from 2010 to 2013 in the Institute of Computing Technology of the Chinese Academy of Sciences. Her research interests include mobile/wireless networks, game theory, crowdsourcing.
\end{IEEEbiography}

\begin{IEEEbiography}[{\includegraphics[width=1in,height=1.25in,clip,keepaspectratio]{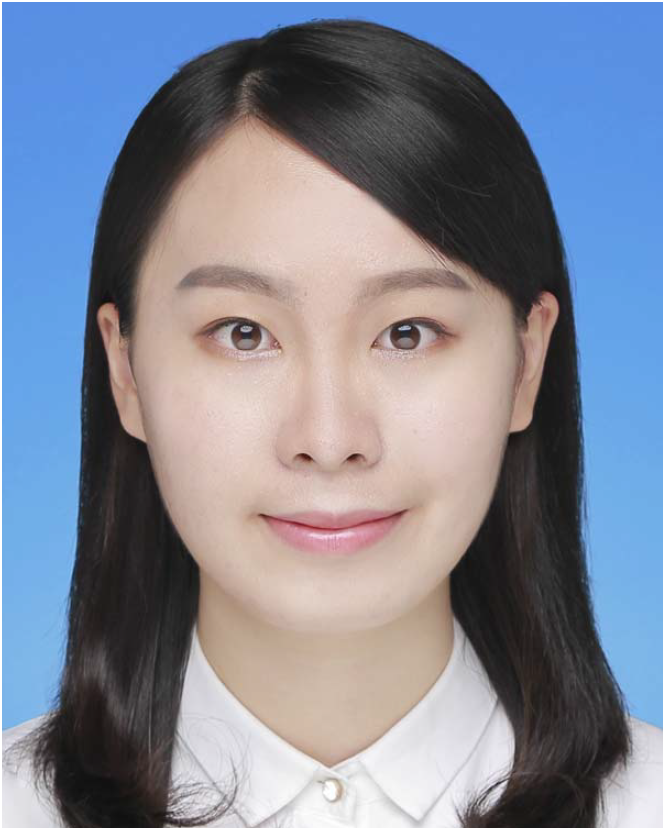}}]{Lina Shi} received the B.S. degree from the School of Artificial Intelligence, Beijing Normal University in 2018. She joined the Center for Big Data Mining \& Knowledge Engineering for research work in September 2018. She is currently a Graduate Student with the School of Artificial Intelligence, Beijing Normal University. Her research interests include game theory and blockchain.
\end{IEEEbiography}

\begin{IEEEbiography}[{\includegraphics[width=1in,height=1.25in,clip,keepaspectratio]{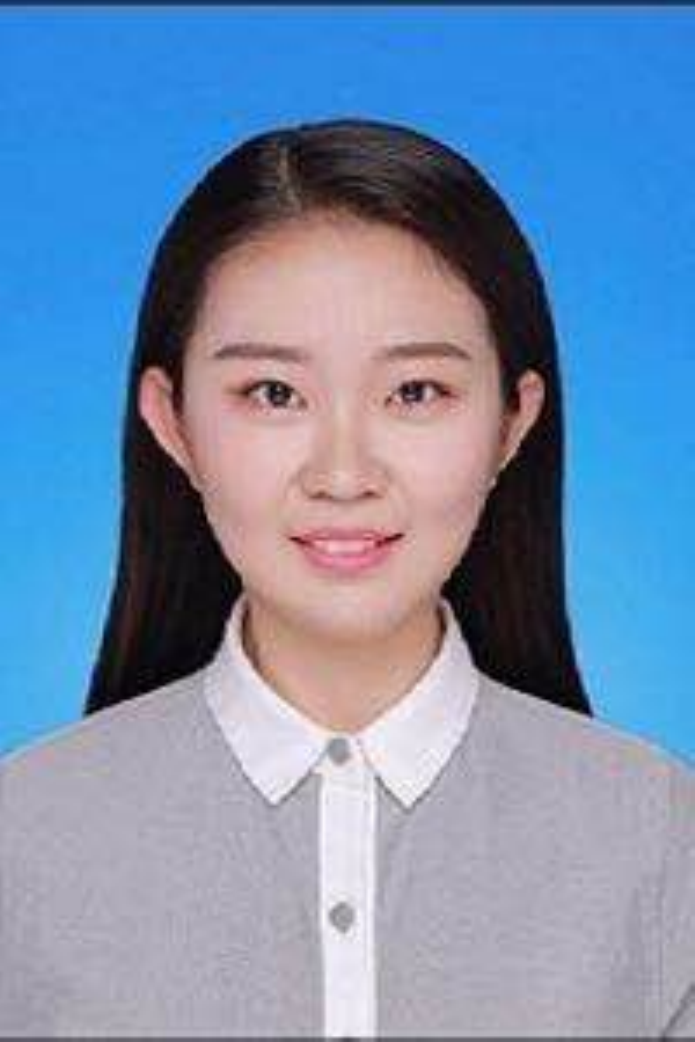}}]
{Hongwei Shi}received her B.S. degree in Computer Science from Beijing Normal University in 2018. Now she is pursuing her Ph.D. degree in Computer Science from Beijing Normal University. Her research interests include blockchain, game theory and combinatorial optimization.
\end{IEEEbiography}

\begin{IEEEbiography}[{\includegraphics[width=1in,height=1.25in,clip,keepaspectratio]{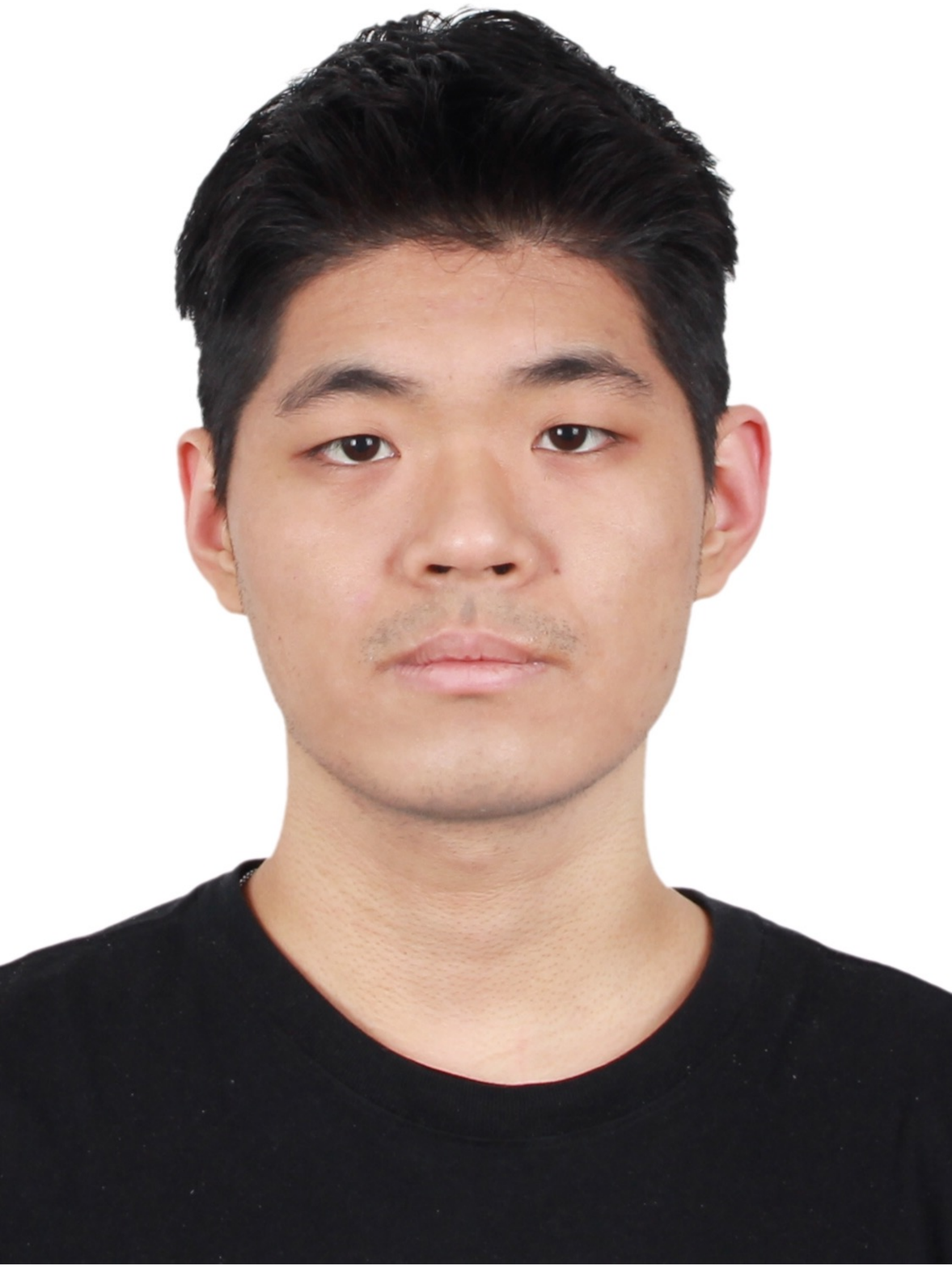}}]
{Yifang Zhang} received the BS degree in network engineering from Beijing Information Science and Technology University, Beijing, China, in 2021. He is currently working toward the graduate degree with the School of Artificial Intelligence, Beijing Normal University, Beijing, China. His research interests include applied cryptography, consensus mechanism, blockchain.
\end{IEEEbiography}

\begin{IEEEbiography}[{\includegraphics[width=1in,height=1.25in,clip,keepaspectratio]{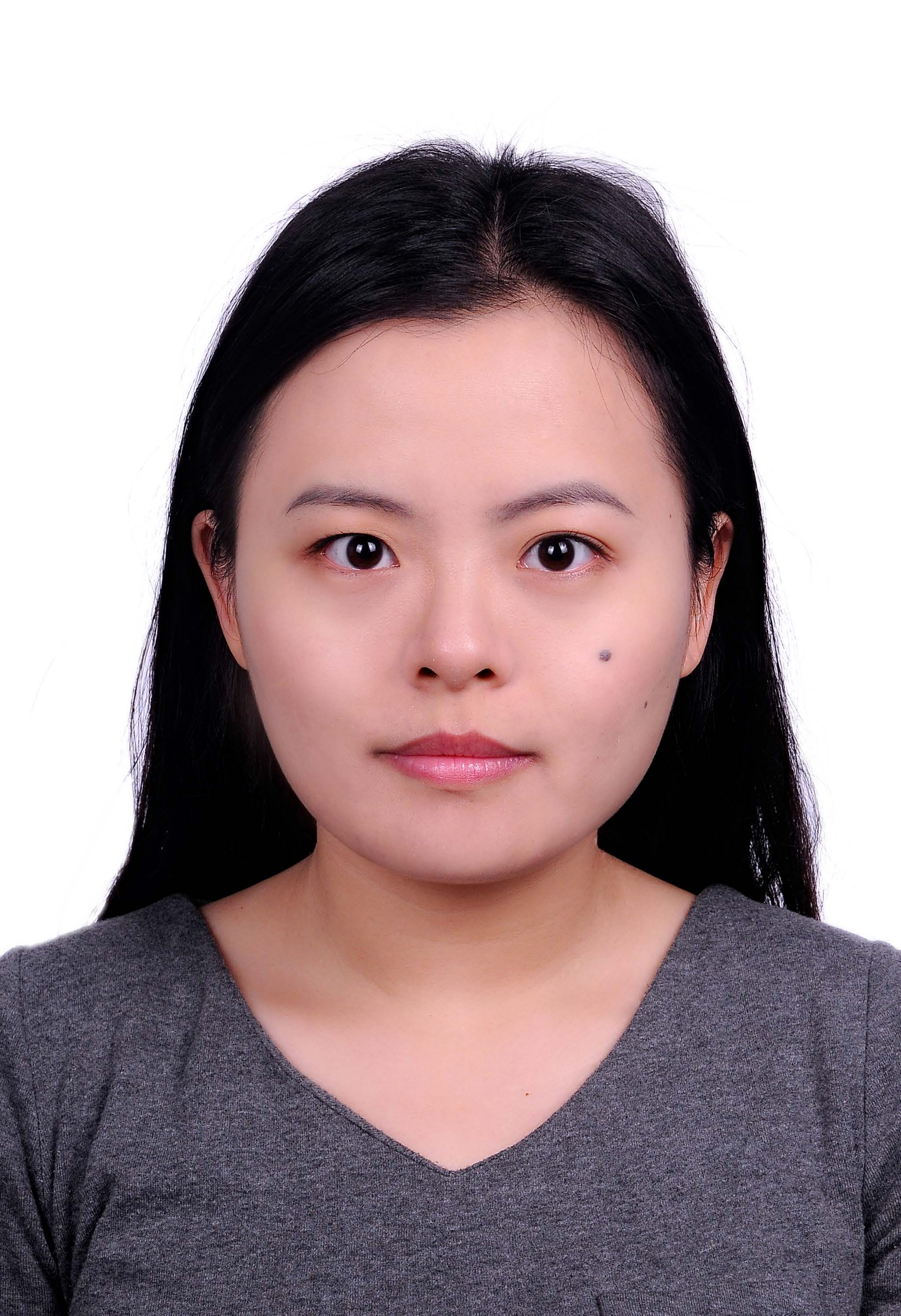}}]
{Qin Hu} received her Ph.D. degree in Computer Science from the George Washington University in 2019. She is currently an Assistant Professor in the department of Computer and Information Science, Indiana University - Purdue University Indianapolis. Her research interests include wireless and mobile security, crowdsourcing/crowdsensing and blockchain.
\end{IEEEbiography}

\begin{IEEEbiography}[{\includegraphics[width=1in,height=1.25in,clip,keepaspectratio]{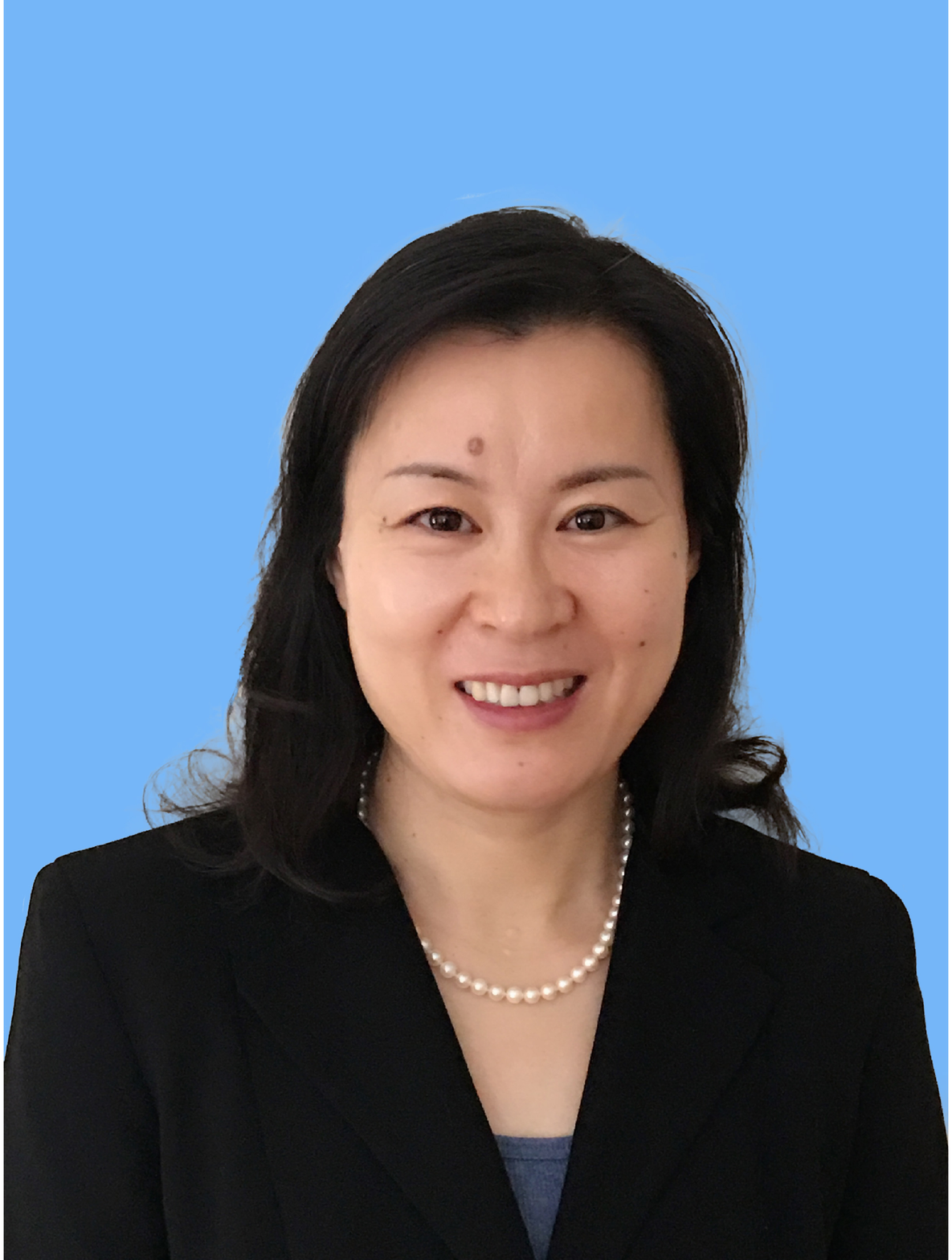}}]{Xiuzhen Cheng} received her M.S. and Ph.D. degrees in computer science from the University of Minnesota Twin Cities in 2000 and 2002, respectively. She is a professor in the School of Computer Science and Technology, Shandong University. Her current research interests focus on privacy-aware computing, wireless and mobile security, dynamic spectrum access, mobile handset networking systems (mobile health and safety), cognitive radio networks, and algorithm design and analysis. She has served on the Editorial Boards of several technical publications and the Technical Program Committees of various professional conferences/workshops. She has also chaired several international conferences. She worked as a program director for the U.S. National Science Foundation (NSF) from April to October 2006 (full time), and from April 2008 to May 2010 (part time). She published more than 170 peer-reviewed papers.
\end{IEEEbiography}
\end{document}